\documentclass[12pt,a4paper]{article}
\usepackage{ucs}
\usepackage[utf8x]{inputenc}
\usepackage[T1]{fontenc}
\usepackage[english]{babel} 
\usepackage{lmodern} 
\usepackage{graphicx}
\usepackage{multirow}
\usepackage[table]{xcolor}
\usepackage{array}
\usepackage{tabularx}
\usepackage{subcaption}
\usepackage{amssymb}
\usepackage[mathcal]{euscript}
\usepackage{geometry}
\usepackage{float}
\usepackage[most]{tcolorbox}
\usepackage{rotating}
\usepackage{tikz}
\usepackage{blindtext}
\usepackage{authblk}
\usepackage{amsmath}
\usepackage[square,numbers]{natbib}
\usepackage[11pt]{moresize}
\usepackage{fixltx2e}
\usepackage{xcolor}
\usepackage{wasysym}
\usepackage{wrapfig}
\usepackage[font=small]{caption}

\begin{document}
\newcommand{\estimates}{\overset{\scriptscriptstyle\wedge}{=}}

\newgeometry{vmargin ={20mm}, hmargin ={18mm,18mm}} 
\title{\textbf{Exchange Bias Demonstrated in Bulk Nanocomposites Processed by High Pressure Torsion}}
\author[1]{M. Zawodzki}
\author[1]{L. Weissitsch}
\author[2]{H. Krenn}
\author[1]{ S. Wurster}
\author[1]{A. Bachmaier}
\affil[1]{Erich Schmid Institute of Materials Science of the Austrian Academy of Sciences, 8700 Leoben, Austria}
\affil[2]{Institute of Physics, University of Graz, 8010 Graz, Austria}
\renewcommand\Affilfont{\itshape\small}
\date{January 2023}
\maketitle
%%\usepackage{showframe} %This line can be used to clearly show the new margins
%%\newgeometry{vmargin ={15mm}, hmargin ={8mm,8mm}}   % set the margins
%%%\authors{M. Zawodzki, L. Weissitsch, H. Krenn, S. Wurster, R. Pippan and A. Bachmaier}
%%\author[1]{M. Zawodzki}
%%\author[1]{L. Weissitsch}
%%\author[2]{H. Krenn}
%%\author[1]{ S. Wurster}
%%\author[1]{R. Pippan}
%%\author[1]{A. Bachmaier}
%%\affil[1]{Erich Schmid Institute of Materials Science of the Austrian Academy of Sciences, 8700 Leoben, Austria}
%%\affil[2]{Institute of Physics, University of Graz, 8010 Graz, Austria}
%%\renewcommand\Affilfont{\itshape\small}
% Abstract (Do not insert blank lines, i.e. \\) 
\abstract{Ferromagnetic (Fe or Fe\textsubscript{20}Ni\textsubscript{80}) and antiferromagnetic (NiO) phases were deformed by high pressure torsion, a severe plastic deformation technique, to manufacture bulk sized nanocomposites and demonstrate an exchange bias, which has been reported predominantly for bilayer thin films. High pressure torsion deformation at elevated temperatures proved to be the key to obtain homogeneous bulk nanocomposites. X-ray diffraction investigations detected nanocrystallinity of the ferromagnetic and antiferromagnetic phases. Furthermore, an additional phase was identified by X-ray diffraction, which formed during deformation at elevated temperatures through the reduction of NiO by Fe. Depending on the initial powder composition of Fe\textsubscript{50}NiO\textsubscript{50} or Fe\textsubscript{10}Ni\textsubscript{40}NiO\textsubscript{50} the new phase was magnetite or maghemite, respectively.  \\
Magnetometry measurements demonstrated an exchange bias in the high pressure torsion processed bulk nanocomposites. Additionally, a tailoring of magnetic parameters was demonstrated by the application of different strains or post-process annealing. A correlation between the amount of applied strain and exchange bias was found. The increase of exchange bias through applied strain was related to the microstructural refinement of the nanocomposite. The nanocrystalline maghemite was considered to have a crucial impact on the observed changes of exchange bias through applied strain.  \\}
\\
% Keywords
\newpage
\section{Introduction}
The introduction and application of nanomaterials is indispensable for progress in technological development. The use of nanomaterials is already well advanced and ranges from applications in medicine (targeted drug delivery for cancer treatment) \cite{Sahu2021}, catalysis (optimisation of the catalysts nanostructure to improve the selectivity and efficiency of catalytic processes \cite{Mitchell2021}, such as the enhanced reduction of volatile organic compounds\cite{Zhang2022}) , gas sensors \cite{Nikolic2020}, microelectronics \cite{Sreenivasulu2022} to engineering \cite{Ovidko2018}. \\
Metallic nanomaterials have outstanding physical properties, such as superior strength, superior hardness and enhanced wear resistance, and attracted the attention of engineers to reduce weight to improve efficiency for transportation vehicles \cite{Valiev2016,Ovidko2018}. \\
The importance of magnetic materials for power efficiency is commonly known. Material scientist focus here on the improvement of the nanostructure of soft- or hard magnetic materials to enhance the magnetic properties \cite{Herzer2013,Coey2011}.The important role of nanocrystalline microstructure can be demonstrated for hard magnetic materials. Here, highly textured nanocrystals are considered to be the key to push the energy product ($BH$) of a hard magnetic material to its theoretical maximum of $BH\textsubscript{max} = 1/4 \mu\textsubscript{0} M\textsubscript{S}\textsuperscript{2}$ ($\mu\textsubscript{0}$...vacuum permeability and M\textsubscript{S}...saturation magnetisation) \cite{Coey2011,Skomski2016}. An increase in $BH$ of the hard magnetic material leads to a reduction of weight and size of permanent magnet motors/generators and this addresses directly the need for power efficiency \cite{Gutfleisch2011}. \\
Although synthesis of nanocrystalline material has been challenging, a variety of methods have been established to obtain the desired nanocrystalline state, (e.g. inert gas condensation, electrodeposition, mechanical alloying, crystallisation out of an amorphous state and severe plastic deformation (SPD)) \cite{Meyers2006}. Regarding synthesis, SPD methods, especially high pressure torsion (HPT), offer unique advantages. For example, it is possible to process a bulk nanocrystalline sample based on powder blends, which allows  a large variety of phase combinations. With HPT it is possible to synthesise supersaturated solid solutions \cite{Kormout2017}  or influence the magnetic properties of the nanocrystalline sample through the application of strain \cite{Wurster2019,Bachmaier2017,Weissitsch2020,Menendez2007}.\\
The focus of this study was on the use of rare-earth free phases due to their abundant availability. Therefore, the choices for a suitable ferromagnetic (FM) material were limited to Fe, Ni or FeNi-alloys. Fe was chosen, because of the vast base of experience available concerning HPT-deformation \cite{Pippan2006}. A suitable alternative FM-phase is the $\gamma$-Fe\textsubscript{20}Ni\textsubscript{80}-alloy, which possesses lower M\textsubscript{S} compared to Fe and far larger domain wall width than Fe or Ni. Those two properties of $\gamma$-Fe\textsubscript{20}Ni\textsubscript{80} are regarded as beneficial to the enhancement of exchange bias (H\textsubscript{eb}). \\
The H\textsubscript{eb} was first observed by Meiklejohn and Bean on Co-CoO nanoparticles and introduced as ‘new unidirectional anisotropy’ \cite{Meiklejohn1956}. H\textsubscript{eb} originates from a FM spin exchange coupling mechanism between the surface spins of adjacent antiferromagnetic (AFM) and FM phases and allows the preservation of an external magnetic field direction through the alignment of AFM surface spins during field cooling (FC) below the Neél temperature (T\textsubscript{N}). This phenomenon causes a biased shift of the hysteresis in the FC direction. \\
A further aim of this study was to use a material, which exhibits AFM properties at room temperature (RT). NiO has the benefit of possessing a face-centred cubic crystal structure with a smaller lattice parameter mismatch to Fe or $\gamma$-Fe\textsubscript{20}Ni\textsubscript{80}-alloy compared to other possible AFM materials (e.g. FeS or $\alpha$-Fe\textsubscript{2}O\textsubscript{3}), and is also AFM far above RT; it was therefore the material of choice.\\
Until recently, such material combinations have mainly been realised by thin film deposition techniques to study the interfacial phenomenon of H\textsubscript{eb} \cite{Blachowicz2021}, due to its technical application in magnetic read heads for hard-disc drives \cite{Childress2005}. Consequently, previous investigations have been limited to 2D structures. A first successful synthesis by HPT of bulk composites, which possess an H\textsubscript{eb}, has been reported for the Co-NiO system \cite{Menendez2007}. \\
The aim in this study was to obtain a bulk nanocomposite possessing enhanced magnetic properties, H\textsubscript{eb} and be able to demonstrate a tailoring of magnetic parameters via applied strain. Furthermore, to investigate in an initial trial the new bulk nanocomposite regarding their nanostructure and the influence of the nanostructure on the magnetic properties.\\
A material synthesis with a metallic and an oxide phase can be a challenging task, especially for HPT, if a homogeneous microstructure is desired. Although oxide ceramics (i.e. Al\textsubscript{2}O\textsubscript{3}, ZrO\textsubscript{2}, TiO\textsubscript{2} and Y\textsubscript{2}O\textsubscript{3}) have been processed previously with HPT \cite{Edalati2019}, the present study has investigated for the first time the deformation process in combination with nanostructural characterisation and magnetic characterisation systematically to provide a detailed description of the FM-AFM bulk nanocomposite.\\
A further aim was to attain deeper insight into the HPT-deformation process of NiO itself and its interaction with the mechanically softer FM-phase. The importance of HPT processing at elevated temperatures was demonstrated to synthesise homogeneous bulk nanocomposites. In addition to a tailoring of magnetic properties, an unusually high Vickers microhardness was detected, when the NiO-phase was preserved during deformation. \\
%%%%%%%%%%%%%%%%%%%%%%%%%%%%%%%%%%%%%%%%%%%
\section{Materials and Methods}
Commercially available powders (Fe - Mateck Fe-99,9\% 100+200 mesh, Ni - Alfa Aesar Ni 99,9\% 100+325 mesh, NiO - Alfa Aesar NiO 99,8 \% 325 mesh) were used. All powder compositions are labelled in –at\%, if not otherwise noted. Powders were stored and prepared inside a glove box filled with Ar-atmosphere. For powder consolidation with HPT an airtight capsule was used to prevent the powder from oxidation. The capsule itself enclosed the HPT-anvils, which can move freely only in axial direction for powder blend compaction. The obtained pellet was used later on in a second step for HPT-deformation Figure ~\ref{figA1}. The processed sample discs had dimensions of ($\varnothing 8  \times 0.6$) $mm$ \cite{Hohenwarter2009}. The process parameters were the following:  applied hydrostatic pressure of 6 $GPa$, rotation speed $\omega  = 1.25$ $min\textsuperscript{-1}$ and deformation temperatures (T\textsubscript{def}) of 200-300°C. The T\textsubscript{def} was provided by inductive heating of both anvils during the processing, and the temperature was controlled by a pyrometer \cite{Pippan2006}. Samples deformed at RT were cooled with pressurised air to ensure temperature stability of the sample. Equivalent von Mises strain ($\epsilon$\textsubscript{vM}) was calculated with \(\epsilon\textsubscript{vM} = (2\pi Nr)/(\sqrt{3}t)\) ($N$...applied rotations, $r$...radius and $t$...sample thickness)\cite{Hohenwarter2009}. \\
Scanning electron microscopy (SEM; LEO-1525 Carl Zeiss GmbH) was used in backscattered electron mode (BSE; Model 7426, Oxford Instruments plc) with an acceleration voltage of 20 $kV$. Energy dispersive X-ray spectroscopy (EDX; XFlash 6–60, Bruker) analysis with SEM was performed at 5 $kV$ to maximise lateral resolution. In order to assign the corresponding radial position to the SEM images, the ends of the examined HPT half discs were determined during the SEM examination. $r$ $\sim$0 $mm$ was defined as the half distance between both ends, and $r$ $\sim$3 $mm$ was determined by the relative distance to $r$ $\sim$0 $mm$. $r$ $\sim$3 $mm$ was cross-checked by measuring the relative distance to the end of the disc, which should be $1$ $mm$. Hardness measurements were performed with a Micromet 5104 from Buehler with an indention load of 500 $g$ along the radial direction every 250 $\mu m$. \\
For X-ray diffraction (XRD), samples were polished on the top side and analysed with a Bruker D2 Phaser (Co-K\textsubscript{$\alpha$} source). XRD data from synchrotron measurements were collected in transmission mode at Deutsches Elektronen-Synchrotron (DESY) at the beamline P07B (high energy materials science) with a beam energy of 87.1 $keV$ and a beam size of $0.5 \times 0.5$ $mm\textsuperscript{2}$. All wide angle X-ray scattering (WAXS) measurements were done in axial direction \cite{Weissitsch2020} and the data were gained with a Perkin Elmer XRD 1621 detector. Peak analysis was done with a self-written script in Octave based on the pseudo-Voigt method to determine integral breadth and coherent scattering domain size (CSDS) \cite{Valiev2000} via Scherrer-relation for spherical crystallites. As  XRD-reference pattern the American Mineralogist Crystal Structure Database (AMCSD) is used (Fe: AMCSD 0012931, Ni: AMCSD 0012932, NiO: AMCSD 0017028, $\gamma$-Fe\textsubscript{2}O\textsubscript{3}: AMCSD 0020516, $\alpha$-Fe\textsubscript{2}O\textsubscript{3}: AMCSD 0000143 and Fe\textsubscript{3}O\textsubscript{4}: AMCSD 0009109). \\
The annealing experiments were done within a vacuum furnace (XTube, type-1200, Xerion Advanced Heating). For sample preparation, a deformed HPT-disc was cut in quarters. One quarter of the sample material was used to obtain probing material for magnetometry measurements from the same radial position, and the other quarters for microstructural characterisation. For the annealing experiment itself, one full quarter and one sample for magnetometry measurements was heated up to 450°C or 550°C in a vacuum of 10\textsuperscript{-6} $mbar$ or better and kept at the specific temperature for one hour. \\
Magnetometry samples were prepared with a diamond wire saw and samples were cut out at desired radial position, having a volume of $\sim$2 $mm$\textsuperscript{3} or less \cite{Weissitsch2020}. Due to the sample dimensions, the applied strain, as mentioned for the magnetometry measurements, is an average value and related to the centre of mass of the measured sample. \\
Field cooling was performed with the applied magnetic field parallel to the radial direction from above the corresponding Néel-temperature from NiO of T\textsubscript{N} $\sim$524 $K$ \cite{Srinivasan1984} to RT inside a vacuum chamber at 10\textsuperscript{-4} $mbar$ or better. An electro magnet provided a magnetic field of $\sim$10 $kOe$. The field cooling was subsequently continued below RT, realised by a superconducting quantum interference device (SQUID) from Quantum Design MPMS-XL-7 from RT to 8 $K$ at 10 $kOe$, which was further used to determine magnetic hysteresis loops for the Fe\textsubscript{10}Ni\textsubscript{40}NiO\textsubscript{50} composition. Measurements were executed at 8 $K$ with a maximum magnetic field strength of $\pm$70 $kOe$ to ensure magnetic saturation. After determining H\textsubscript{C1} and  H\textsubscript{C2}, the H\textsubscript{eb} can be calculated with $H\textsubscript{eb} = (H\textsubscript{C1}+H\textsubscript{C2})/2$. The coercivity of the symmetric hysteresis is determined by $H\textsubscript{C} = ((-)H\textsubscript{C1}+H\textsubscript{C2})/2$. M\textsubscript{S} was approximated with a linear fit from the saturated branches of hysteresis to obtain M\textsubscript{S} at the intercept with the ordinate. \\
%%%%%%%%%%%%%%%%%%%%%%%%%%%%%%%%%%%%%%%%%%
\section{Results}
\subsection{Hardness and microstructural evolution}
\subsubsection{Fe\textsubscript{50}NiO\textsubscript{50} composition}
\begin{figure}[H]
\center
\includegraphics[width=16.5 cm]{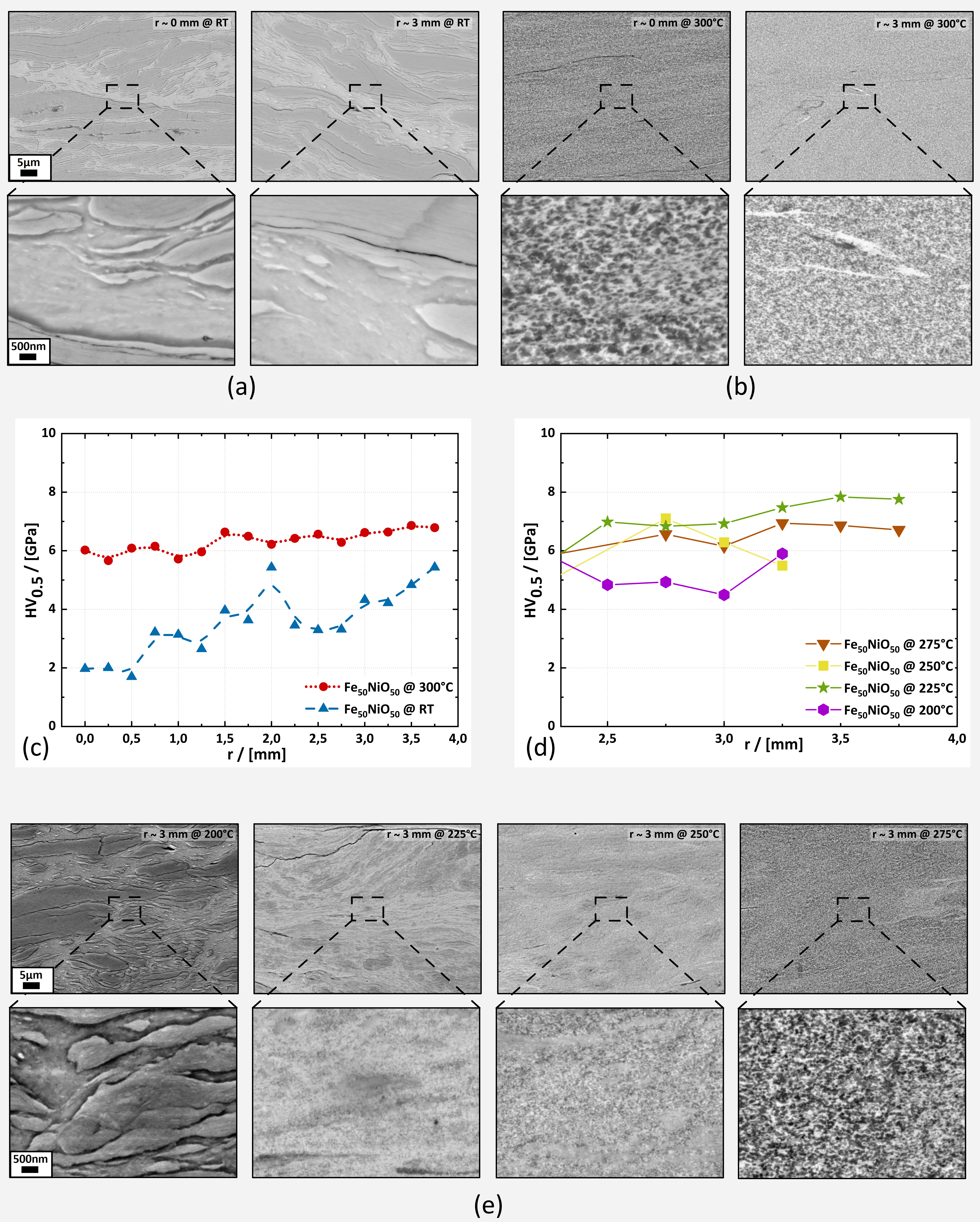}
\caption{SEM micrographs in BSE mode depict samples deformed at RT (\textbf{a}) and 300°C (\textbf{b}) at $r = 0$ $mm$ and $r = 3$ $mm$. EDX spectroscopy identified the dark phase as NiO for (a) or Fe\textsubscript{3}O\textsubscript{4} for (b). (\textbf{c}) Vickers microhardness of Fe\textsubscript{50}NiO\textsubscript{50} samples along radial direction.The sample processed at 300°C had a higher saturation microhardness compared to the sample deformed at RT. (\textbf{d}) Vickers microhardness values form  $r = 2.5$ $mm$ to  $r = 3.75$ $mm$. Higher deformation temperatures led to a homogeneous microstructure and therefore higher hardness values. (\textbf{e}) SEM micrographs in BSE mode gained of Fe\textsubscript{50}NiO\textsubscript{50} samples at the outer radii ($r$ $\sim$3 $mm$). An onset of homogenisation of microstructure could be seen at 225°C and gradually improved towards 275°C. The micron bar applies to all micrographs in the same row. \label{fig1}}
\end{figure}

\begin{figure}[H]
 \center
\includegraphics[width=15 cm]{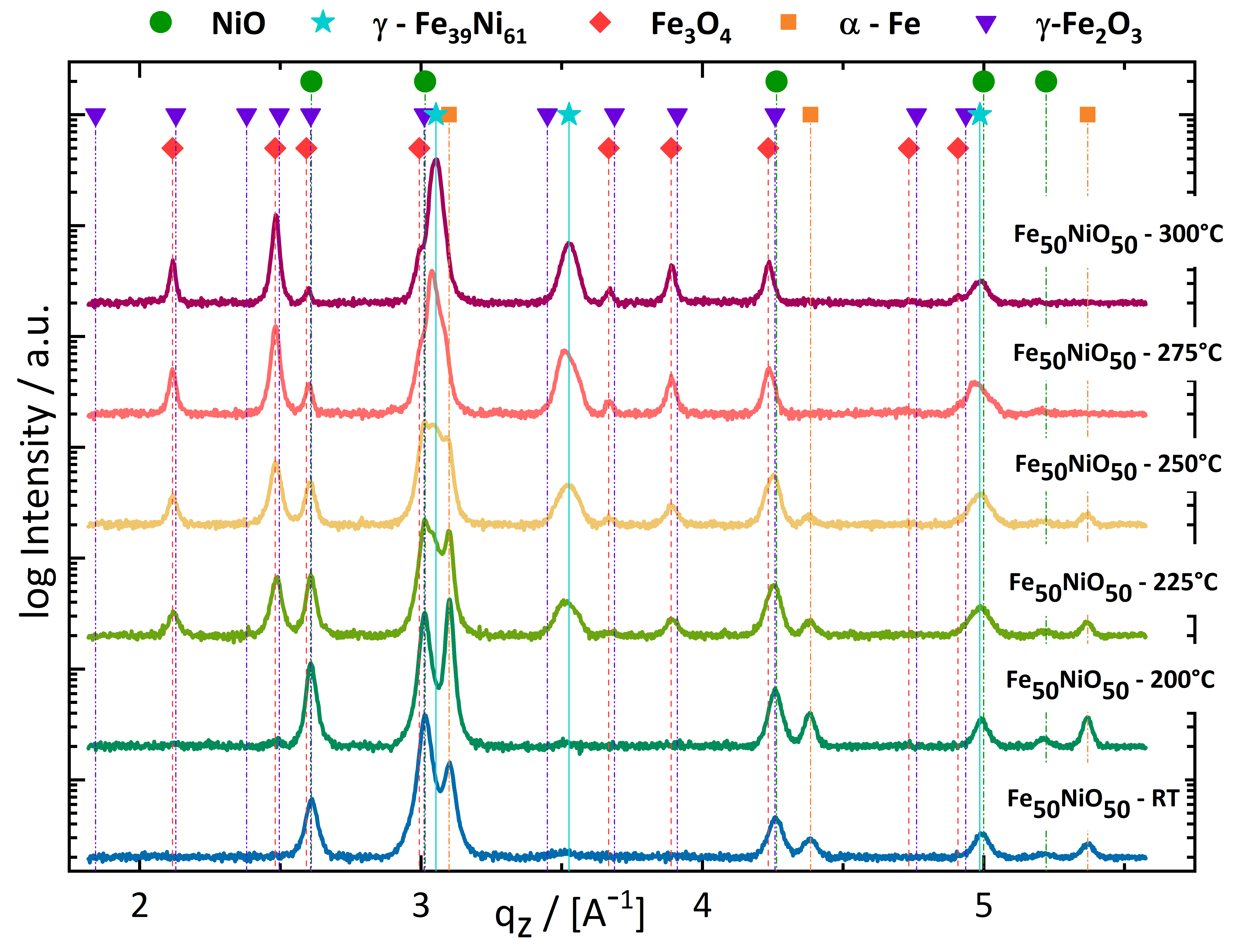}
\caption{XRD-data is displayed of the Fe\textsubscript{50}NiO\textsubscript{50} sample, which was measured with laboratory equipment. The formation of Fe\textsubscript{3}O\textsubscript{4}, through the reduction of NiO by Fe, was seen for samples deformed at T\textsubscript{def} $\geq$225°C. $\alpha$-Fe was not detectable at T\textsubscript{def} $\geq$275°C. \label{fig2}}
\end{figure}

The first composition investigated was the binary Fe\textsubscript{50}NiO\textsubscript{50} (Fe\textsubscript{42.8}NiO\textsubscript{57.2} in -wt\%) blend. Figures ~\ref{fig1}a and b present SEM micrographs at $r$ $\sim$0 $mm$ and $r$ $\sim$3 $mm$. Both samples were processed with 100 rotations and deformation temperature was increased from RT (Figure ~\ref{fig1}a) to 300°C (Figure ~\ref{fig1}b). This yielded an applied strain for T\textsubscript{def} = RT of $\epsilon$\textsubscript{vM} $\sim$1760 and for T\textsubscript{def} = 300°C  a $\epsilon$\textsubscript{vM} $\sim$1490 at $r$ $\sim$3 $mm$, due to a slight difference in height. For the sample processed at RT, EDX measurements confirmed that the dark phase seen in BSE-contrast consisted of NiO, while the bright phase was identified as Fe-phase. \\
When processed at RT, both phases formed lamella structures (Figure ~\ref{fig1}a). Although the applied strain was high, the expected refinement of the two-phase microstructure was not observed. Within the phases, a grain refinement was evident  at $r$ $\sim$3 $mm$. Vickers microhardness measurements detected an increase of microhardness, (Figure ~\ref{fig1}c) and confirmed the ongoing refinement, which was seen in SEM within the phases. This observation indicated that the applied strain had acted in the composite to a certain degree. However, whether the total amount of applied strain had been absorbed by the composite through deformation was in question. The deformation at RT of samples with NiO of 50-at\% (Figure ~\ref{fig1}a) or higher was considered to be unsuccessful because the samples had inhomogeneous microstructure and the FM-phase dimensions were too large for observing an H\textsubscript{eb}. \\
Deformation at elevated temperatures improved the deformation behaviour of the nanocomposites and led to a homogeneous microstructure (Figure ~\ref{fig1}b). SEM micrographs indicated that a homogeneous two-phase microstructure had been formed, having microstructural refinement from $r$ $\sim$0 $mm$ to $r$ $\sim$3 $mm$. The large NiO structures, which were observed for T\textsubscript{def} = RT, were not observed for T\textsubscript{def} = 300°C in the SEM micrographs along the sample radius. Here, the dark phase, which is shown in BSE-contrast, was determined by EDX measurements in combination with XRD-results (Figure ~\ref{fig2}) as Fe\textsubscript{3}O\textsubscript{4} and the bright phase was identified as  a FeNi-phase.\\
Vickers microhardness measurements detected a large difference in hardness between samples deformed at RT and 300°C (Figure ~\ref{fig1}c). For the T\textsubscript{def} = 300°C composite, hardness values were considerably higher than for the RT sample, varying from 6  $GPa$ at $r$ $\sim$0 $mm$ ($\epsilon$\textsubscript{vM} $\sim$0) to 6.8 $GPa$ at $r$ $\sim$3 $mm$ ($\epsilon$\textsubscript{vM} $\sim$1740). Above $r$ $\sim$1.5 $mm$ the T\textsubscript{def} = 300°C composite became microstructurally saturated according to Vickers microhardness measurements.\\
%%%
Further deformation experiments were conducted at deformation temperatures between RT and 300°C (T\textsubscript{def} = 200°C, 225°C, 250°C and 275°C) to gain insight into the onset of the microstructural homogenisation. Figure ~\ref{fig1}e displays SEM micrographs at $r$ $\sim$3 $mm$. At T\textsubscript{def} = 200°C, the microstructure was identical to that at RT; however, at T\textsubscript{def} = 225°C, the large NiO structures had become more and more refined through the application of strain. This process continued for T\textsubscript{def} = 250°C and T\textsubscript{def} = 275°C, gradually extending from outer radius to inner radius. At T\textsubscript{def} = 275°C, the microstructure is comparable to the T\textsubscript{def} = 300°C samples at $r$ $\sim$3 $mm$, but still large NiO structures were observed for T\textsubscript{def} = 275°C at $r \leq2$ $mm$ (not shown).\\ 
The Vicker microhardness values that were obtained, reflected the observations from SEM. In Figure ~\ref{fig1}d hardness results at the outer radius are summarised. The inhomogeneous microstructure of the sample deformed at T\textsubscript{def} = 200°C had a lower microhardness (5 $GPa$) than those of samples deformed at temperatures T\textsubscript{def} $\geq$225°C, which were approximately 6 to 7 $GPa$ at  $r$ $\sim$3 $mm$ and similar to the hardness values of the sample deformed at T\textsubscript{def} = 300°C. All samples processed at elevated temperatures exhibited significantly higher Vickers microhardness values than the sample processed at RT. \\
Phase analysis with XRD detected Fe and NiO phases for the samples deformed at RT and 200°C  with an onset of Fe\textsubscript{3}O\textsubscript{4} formation at T\textsubscript{def} =200°C (Figure ~\ref{fig2}). An increase of deformation temperature from T\textsubscript{def} = 225°C to T\textsubscript{def} = 300°C yielded a continuously growing amount of Fe\textsubscript{3}O\textsubscript{4}, caused by the reduction of NiO through Fe. 
Ni became available to build the $\gamma$-FeNi phase, which was detected in the XRD-scans. The content of Fe within the $\gamma$-FeNi phase could be approximated through the shift to lower q-values. This shift is caused by the substitution of Fe-atoms within the Ni lattice, resulting in an enlargement of lattice parameter. This enlargement is caused by the higher atomic radius of Fe compared to Ni \cite{Owen1937}. The shifted peaks of $\gamma$-FeNi phase for T\textsubscript{def} = 300°C implied a composition of approximately $\gamma$-Fe\textsubscript{39}Ni\textsubscript{61}.\\
The quality of the XRD patterns and similarity of Fe\textsubscript{3}O\textsubscript{4} and maghemite ($\gamma$-Fe\textsubscript{2}O\textsubscript{3}) patterns allowed a determination of Fe\textsubscript{x}O\textsubscript{y} phase only at higher q\textsubscript{z} $\geq$3.5  \AA\textsuperscript{-1}. The peaks at q\textsubscript{z} = 3.889 \AA\textsuperscript{-1} and  q\textsubscript{z} = 4.235 \AA\textsuperscript{-1}, for example, are in accordance with the Fe\textsubscript{3}O\textsubscript{4} pattern. Presence of  $\gamma$-Fe\textsubscript{2}O\textsubscript{3} could not be excluded, but apparently Fe\textsubscript{3}O\textsubscript{4} dominated the XRD-pattern; $\gamma$-Fe\textsubscript{2}O\textsubscript{3} was therefore omitted in the following consideration for the Fe\textsubscript{50}NiO\textsubscript{50} nanocomposite.\\
The assumption that the $\alpha$-Fe phase had been completely consumed by the formation of Fe\textsubscript{3}O\textsubscript{4} or $\gamma$-FeNi, allowed estimation of the residual NiO. Using the known composition of $\gamma$-Fe\textsubscript{39}Ni\textsubscript{61} phase for the sample deformed at T\textsubscript{def} = 300°C,  yielded a residual NiO-phase of maximum 16-wt\%, which should have been left after synthesis inside the sample, but could not be resolved by the laboratory XRD-equipment due to overlapping XRD-peaks (Figure ~\ref{fig2}). Although the sample deformed at T\textsubscript{def} = 300°C had a promising microstructure, this nanocomposite was considered unsuitable for observing a significant H\textsubscript{eb} in subsequent magnetometry measurements because it did not contain large amounts NiO according to the XRD results.  \\
%%%
The CSDS was calculated via Scherrer-relation of Fe\textsubscript{50}NiO\textsubscript{50} sample at RT and T\textsubscript{def} = 300°C displayed nanocrystallinity, whereby Fe\textsubscript{3}O\textsubscript{4} had the largest CSDS. Considering residual strain with the Williamson-Hall (WH) method \cite{Valiev2000} and comparing results, strain in the Fe\textsubscript{3}O\textsubscript{4}-phase is almost half compared to $\gamma$-FeNi. Strain-free CSDS estimations yielded approximately 45 $nm$ for both analysed phases at T\textsubscript{def} = 300°C (Table~\ref{tab1}). \\
%%%
\begin{table}[H]
\caption{Summary of peak analysis executed for Fe\textsubscript{50}NiO\textsubscript{50} sample deformed at RT and 300°C. <D\textsubscript{x}> is the average crystallite size or CSDS, D\textsubscript{x-WH} represents the strain-less CSDS result and $\epsilon$\textsubscript{x-WH} residual strain in the crystallite, both values are obtained via the WH method.\label{tab1}}
	
		\newcolumntype{C}{>{\centering\arraybackslash}X}
		\begin{tabularx}{1\textwidth}{CCCCCCC}
			\hline\hline
			\empty	& \textbf{<D\textsubscript{Fe}> / [$nm$]}	& \textbf{D\textsubscript{Fe-WH} / [$nm$]}     & \textbf{$\epsilon$\textsubscript{Fe-WH}} & \textbf{<D\textsubscript{NiO}> / [$nm$]} & \textbf{D\textsubscript{NiO-WH} / [$nm$]} & \textbf{$\epsilon$\textsubscript{NiO-WH}}\\
			\hline
		 \empty	& \empty		&\empty	&\empty	&\empty	&\empty	& \empty\\
\multirow[m]{1}{*}{Fe\textsubscript{50}NiO\textsubscript{50}}	&11$\pm$2&		40&	2$\cdot$10$^{-2}$ &11$\pm$2	&30	&		2$\cdot$10$^{-2}$	\\
			  	     @T\textsubscript{def} =RT              & \empty			&\empty		&\empty &\empty &\empty	& \empty\\
			             	     
\hline
			\empty	& \textbf{<D\textsubscript{FeNi}> / [$nm$]}	& \textbf{D\textsubscript{FeNi-WH} / [$nm$]}     & \textbf{$\epsilon$\textsubscript{FeNi-WH}} & \textbf{<D\textsubscript{Fe3O4}> / [$nm$]} & \textbf{D\textsubscript{Fe3O4-WH} / [$nm$]} & \textbf{$\epsilon$\textsubscript{Fe3O4-WH}}\\			             	      
			  \hline
			  \empty	& \empty	&\empty	&\empty	&\empty	&\empty		& \empty\\
\multirow[m]{1}{*}{Fe\textsubscript{50}NiO\textsubscript{50} }    &12$\pm$2&	45&	3$\cdot$10$^{-2}$ 	&19$\pm$7 &43	&	1$\cdot$10$^{-2}$ \\
			@T\textsubscript{def}=300°C  	                 & \empty			& \empty			& \empty\\

			\hline\hline
		\end{tabularx}
	
	%\noindent{\footnotesize{* Tables may have a footer.}}
\end{table}
\subsubsection{Fe\textsubscript{10}Ni\textsubscript{40}NiO\textsubscript{50} composition}
\begin{figure}[H]
\center
\includegraphics[width=18 cm]{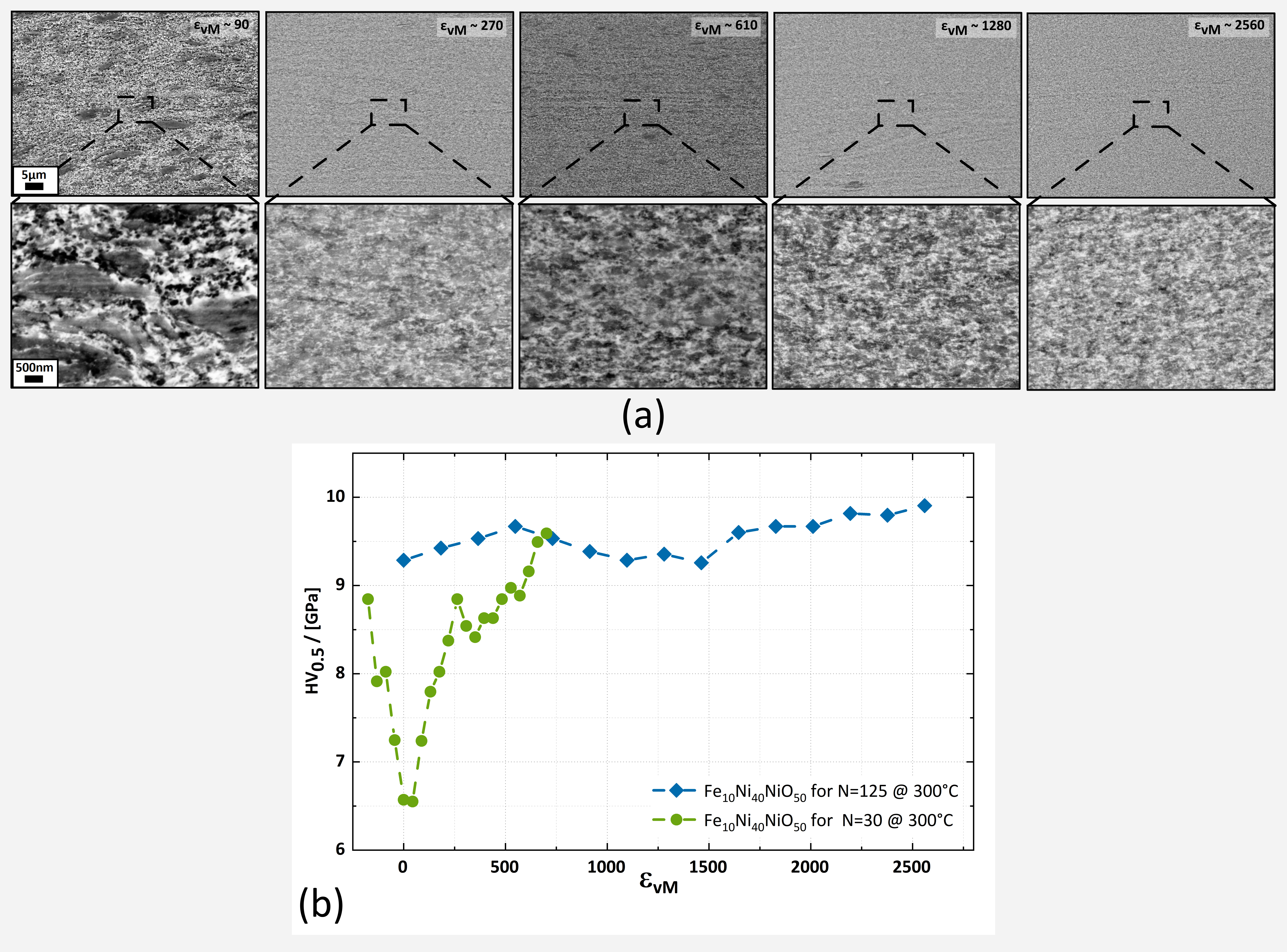}
\caption{ (\textbf{a}) BSE micrographs of the Fe\textsubscript{10}Ni\textsubscript{40}NiO\textsubscript{50} samples. Microstructural refinement was detected from left to right. Two samples were deformed to obtain the depicted micrographs. First samples was deformed with 30 rotations and micrographs of an applied strain of $\epsilon\textsubscript{vM}$ $\sim$90 to $\epsilon\textsubscript{vM}$ $\sim$610 were taken. Second samples was deformed with 125 rotations and micrographs of an applied strain of $\epsilon\textsubscript{vM}$ $\sim$270, $\epsilon\textsubscript{vM}$ $\sim$1280 and $\epsilon\textsubscript{vM}$ $\sim$2560 were taken. The micron bar applies to all micrographs in the same row. (\textbf{b}) Vickers microhardness of both Fe\textsubscript{10}Ni\textsubscript{40}NiO\textsubscript{50} samples deformed at 300°C. \label{fig3}}
\end{figure}

\begin{figure}[H]

\center
\includegraphics[width=15 cm]{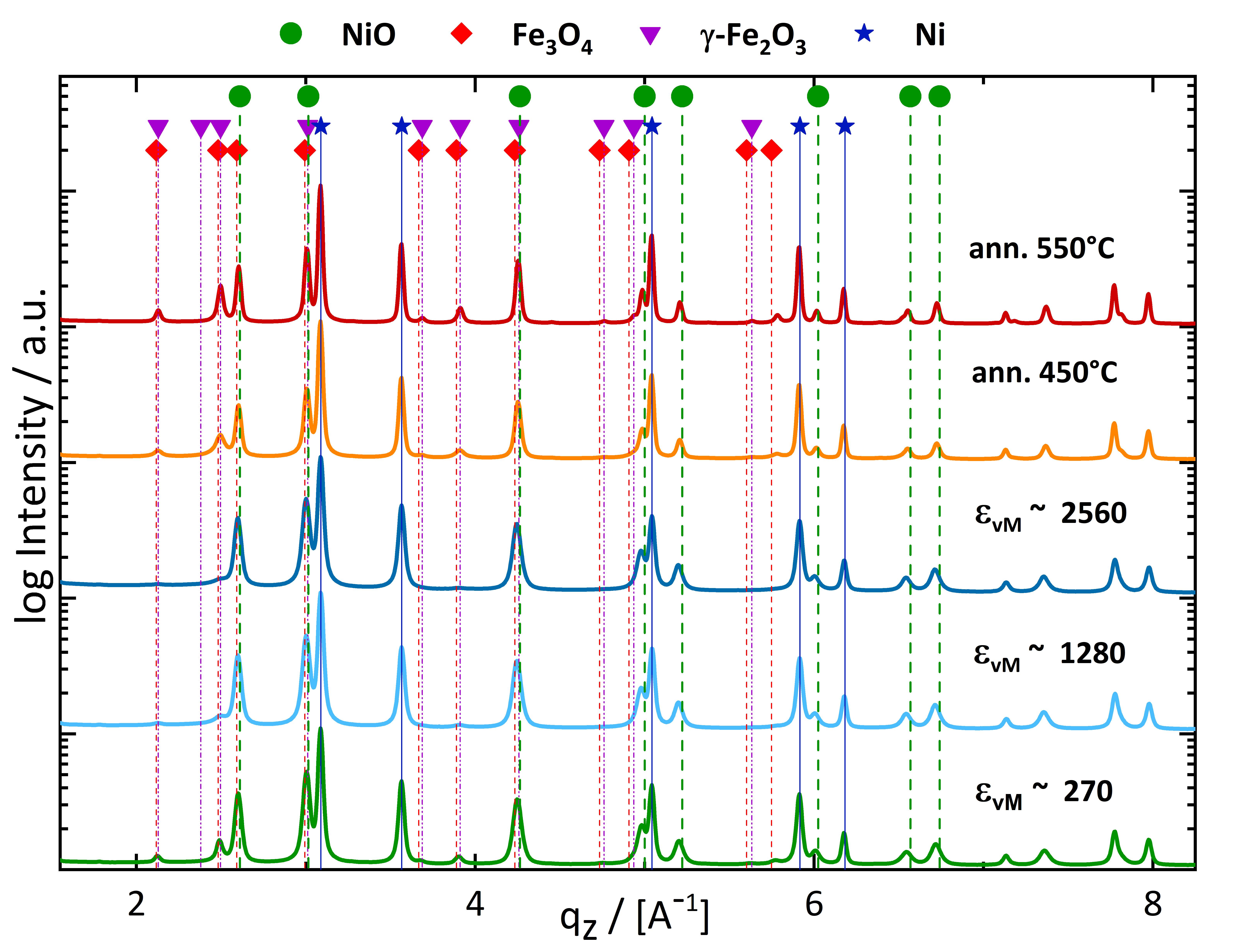}

\caption{ Synchrotron WAXS results are shown for different applied strains of the Fe\textsubscript{10}Ni\textsubscript{40}NiO\textsubscript{50} sample. Starting from bottom:  $\epsilon\textsubscript{vM}$ $\sim$270, $\epsilon\textsubscript{vM}$ $\sim$1280 and $\epsilon\textsubscript{vM}$ $\sim$2560. The two annealed samples are at the top: atop the sample annealed at 550°C and below the sample annealed at 450°C. At low q-values the development of $\gamma$-Fe\textsubscript{2}O\textsubscript{3} peaks were visible. $\gamma$-Fe\textsubscript{2}O\textsubscript{3}-peaks became weaker with high applied strains due to a reduction of CSDS. \label{fig4}}
\end{figure}

To reduce the loss of NiO at 300°C, another composition of Fe\textsubscript{10}Ni\textsubscript{40}NiO\textsubscript{50} (Fe\textsubscript{8.4}Ni\textsubscript{35.4}NiO\textsubscript{56.2} in wt\%)  was tested with the idea to conserve the NiO-phase by alloying Ni initially with Fe to prevent the reduction of NiO. \\
The microstructures depicted in Figure ~\ref{fig3}a were collected from two samples; one has been deformed with 30 rotations the other with 125 rotations; all other parameters were equal. For the micrographs shown, the applied strain was $\epsilon\textsubscript{vM}$ $\sim$90 ($\triangleq$ $r$ $\sim$0.5 $mm$) and $\epsilon\textsubscript{vM}$ $\sim$610 ($\triangleq$ $r$ $\sim$3.5 $mm$) for the sample with 30 rotations. For the sample with 125 rotations, the applied strain was equivalent to: $\epsilon\textsubscript{vM}$ $\sim$270 ($\triangleq$ $r$ $\sim$0.375 $mm$), $\epsilon\textsubscript{vM}$ $\sim$1280 ($\triangleq$ $r$ $\sim$1.75 $mm$) and $\epsilon\textsubscript{vM}$ $\sim$2560 ($\triangleq$ $r$ $\sim$3.5 $mm$). Due to the homogeneous microstructure, a uniform deformation of the sample material was assumed and applied strain was therefore used rather than the radius. The dark phase was identified as NiO and the bright phase as Ni. The microstructure was different from the Fe\textsubscript{50}NiO\textsubscript{50} sample deformed at T\textsubscript{def} = 300°C (for comparison Figure ~\ref{fig1}b). Through the preservation of NiO, a fragmentation process of NiO was realised, leading to a highly diverse Ni and NiO nanocomposite structure. Additionally, SEM micrographs demonstrated a refinement of the nanocomposite microstructure with an increase of applied strain. \\
Vickers microhardness measurements along the radial direction demonstrated a microstructural saturation at $\epsilon\textsubscript{vM}$ $\sim$700 of 9.3 $GPa$ in microhardness for the Fe\textsubscript{10}Ni\textsubscript{40}NiO\textsubscript{50} sample after 125 rotations, (Figure ~\ref{fig3}b). Beyond $\epsilon\textsubscript{vM}$ $\sim$700, further deformation caused only a small increase in microhardness of approximately 0.6 $GPa$ to 9.9 $GPa$, which correlated with the minor refinement seen in SEM between $\epsilon\textsubscript{vM}$ $\sim$1280 to $\epsilon\textsubscript{vM}$ $\sim$2560. Vickers microhardness values of 9.3 to 9.9 $GPa$ were significantly higher than previously reported results from similar nanocrystalline materials \cite{Siow2004,Bachmaier2009}.\\
%%%                                                       
Synchrotron WAXS investigations were conducted along the axial direction for following applied strains: $\epsilon\textsubscript{vM}$ $\sim$270, $\epsilon\textsubscript{vM}$ $\sim$1280, and $\epsilon\textsubscript{vM}$ $\sim$2560. Detected phases were NiO, Ni and $\gamma$-Fe\textsubscript{2}O\textsubscript{3}. However, the $\gamma$-Fe\textsubscript{2}O\textsubscript{3} peaks could not be directly identified. The comparison of WAXS patterns of the as-deformed and post-process annealed samples (Section \ref{section:Annealed}) showed a slight shift from the Fe\textsubscript{3}O\textsubscript{4} to the $\gamma$-Fe\textsubscript{2}O\textsubscript{3} reference pattern (Figure ~\ref{fig4}). This shift became more evident through regarding the calculated peak centres.\\
According to the WAXS measurements, the FM-phase consisted of solely Ni. An expected substitution of Fe was thus not detectable with WAXS and implied that the main part of Fe was consumed by the formation of $\gamma$-Fe\textsubscript{2}O\textsubscript{3}. The WAXS results allowed estimation of residual amount of NiO within sample and yielded the following composition: Ni\textsubscript{48.6}NiO\textsubscript{39.4} and ($\gamma$-Fe\textsubscript{2}O\textsubscript{3})\textsubscript{12}, all in wt\%. \\    
Integral peak breadth analysation with the Scherrer-relation showed minor changes of CSDS for Ni and NiO, but the CSDS of $\gamma$-Fe\textsubscript{2}O\textsubscript{3} reduced by 50\% with applied strain (Table ~\ref{tab2}). Residual strain was approximated via the WH method and for NiO was approximately a factor of 10 higher than for Ni. \\
\unskip   
\begin{table}[H]
\caption{A summary of integral peak breadth analysis is shown, which was done for the Fe\textsubscript{10}Ni\textsubscript{40}NiO\textsubscript{50} sample deformed at T\textsubscript{def} = 300°C. <D\textsubscript{x}> is the average crystallite size or CSDS, D\textsubscript{x-WH} represents the strain-less CSDS result and $\epsilon$\textsubscript{x-WH} residual strain in the crystallite both values were obtained by the WH method.\label{tab2}}
	
		\newcolumntype{C}{>{\centering\arraybackslash}X}
		\begin{tabularx}{1\textwidth}{CCCCCCCC}
			\hline\hline
			\empty	& \textbf{<D\textsubscript{Ni}> / [$nm$]}	& \textbf{D\textsubscript{Ni-WH} / [$nm$]}     & \textbf{$\epsilon$\textsubscript{Ni-WH}} & \textbf{<D\textsubscript{NiO}> / [$nm$]} & \textbf{D\textsubscript{NiO-WH} / [$nm$]} & \textbf{$\epsilon$\textsubscript{NiO-WH}} &<\textbf{D\textsubscript{$\gamma$-Fe2O3}}> / [$nm$]\\
			\hline
$\epsilon$\textsubscript{vM}$\sim$270&	18$\pm$2&		21&	3$\cdot$10$^{-3}$ &11$\pm$3	&25	&		2$\cdot$10$^{-2}$ &	 12$\pm$2\\
$\epsilon$\textsubscript{vM}$\sim$1280&	15$\pm$5&	19&	2$\cdot$10$^{-3}$&	11$\pm$2&	20&		1$\cdot$10$^{-2}$ & 9$\pm$4\\
$\epsilon$\textsubscript{vM}$\sim$2560&	15$\pm$4&	18&	2$\cdot$10$^{-3}$&	11$\pm$2&	19&		1$\cdot$10$^{-2}$ & 6$\pm$2\\
\hline

ann.450°C&		21$\pm$2&	22&	1$\cdot$10$^{-3}$&	14$\pm$3&	22&			9$\cdot$10$^{-3}$& 8$\pm$2\\
ann.550°C&		24$\pm$2&	25&	0.5$\cdot$10$^{-3}$&		17$\pm$3&	26&			7$\cdot$10$^{-3}$&14$\pm$2\\

			\hline\hline
		\end{tabularx}
	
	%\noindent{\footnotesize{* Tables may have a footer.}}
\end{table}
%%%
\subsubsection{Annealing of Fe\textsubscript{10}Ni\textsubscript{40}NiO\textsubscript{50} composition}
\label{section:Annealed}
\begin{figure}[H]
\center
\includegraphics[width=16 cm]{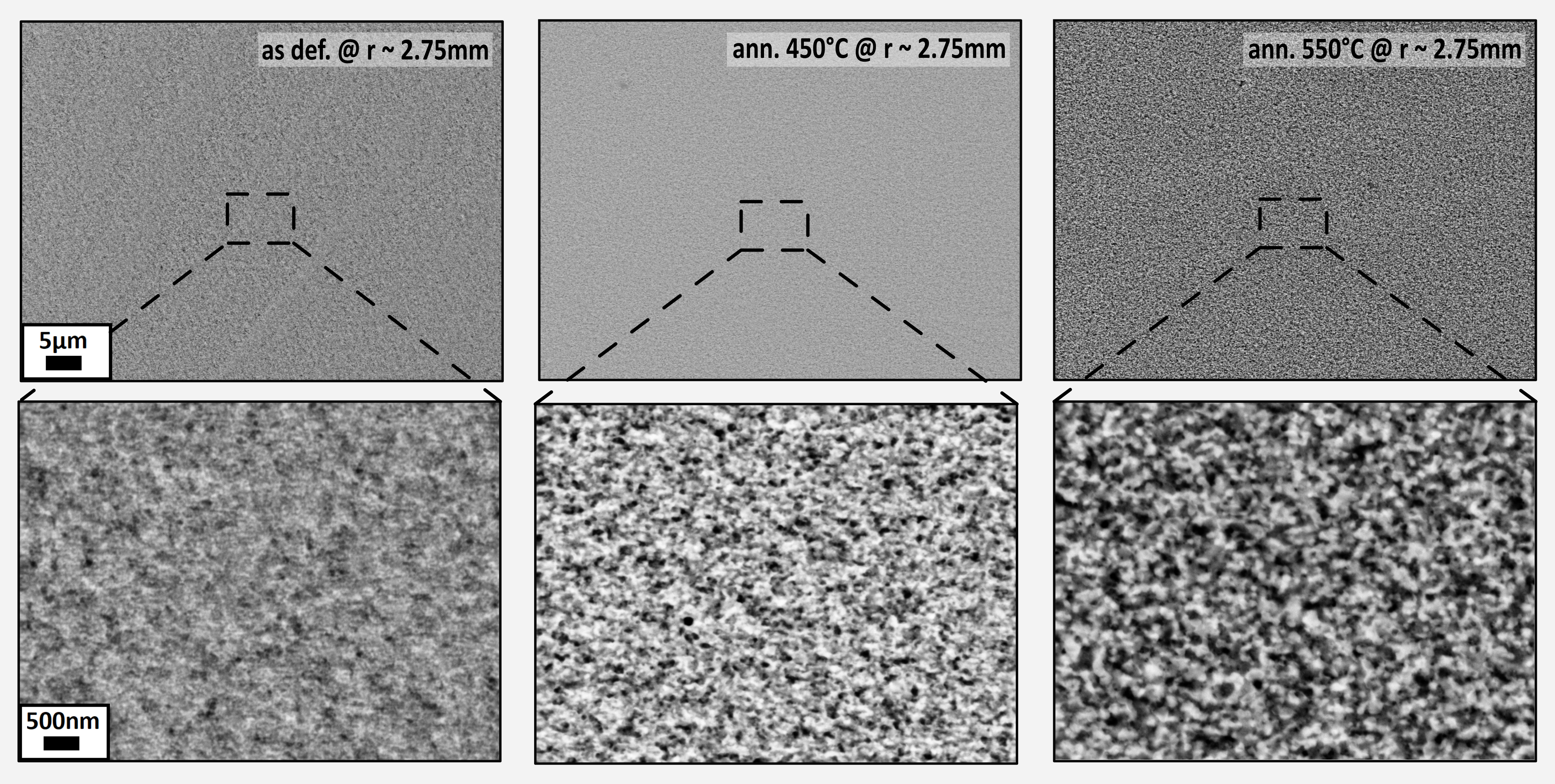}
\caption{ SEM micrographs were done in BSE mode of as-deformed and annealed samples of the Fe\textsubscript{10}Ni\textsubscript{40}NiO\textsubscript{50} composition. The annealing initiated a phase growth and change of interface morphology for the samples annealed at 450°C and 550°C. The micron bar applies to all micrographs in the same row. \label{fig5}}
\end{figure}

Because of the complex microstructure of the nanocomposite after deformation the effects of post-process annealing were investigated to understand the influence of phase growth and phase interface morphology on the magnetic properties of the Fe\textsubscript{10}Ni\textsubscript{40}NiO\textsubscript{50} sample better. Hence, sample material with a large amount of applied strain $\epsilon\textsubscript{vM}$ $\sim$1150 ($r$ $\sim$2.75 $mm$) was used to have . \\
The SEM micrographs are displayed in Figure ~\ref{fig5} for the following states: as-deformed, annealed in vacuum at 450°C and 550°C. For the as-deformed state, the complex phase interface was assumed to have large variations in topography at a nanometre scale, which was expected to simplify through annealing. The contrast in BSE mode has been enhanced for the annealed samples. This enhancement indicated a phase growth of the bright and dark phases. The annealing led to a reduction of phase interface, which was expected to be predominantly driven by the optimization of Gibbs free energy. \\
XRD-peak analysation affirmed the rise of CSDS for all three detected phases (Table ~\ref{tab2}), whereby $\gamma$-Fe\textsubscript{2}O\textsubscript{3} had the highest increase in relative crystallite size of approximately two times to its initial CSDS in the as-deformed state. The relative CSDS growth of Ni and NiO-phase during annealing was significantly smaller.\\
%%%
%%%
\subsection{Magnetic characterisation}
\begin{figure}[H]
\center
\includegraphics[width=17 cm]{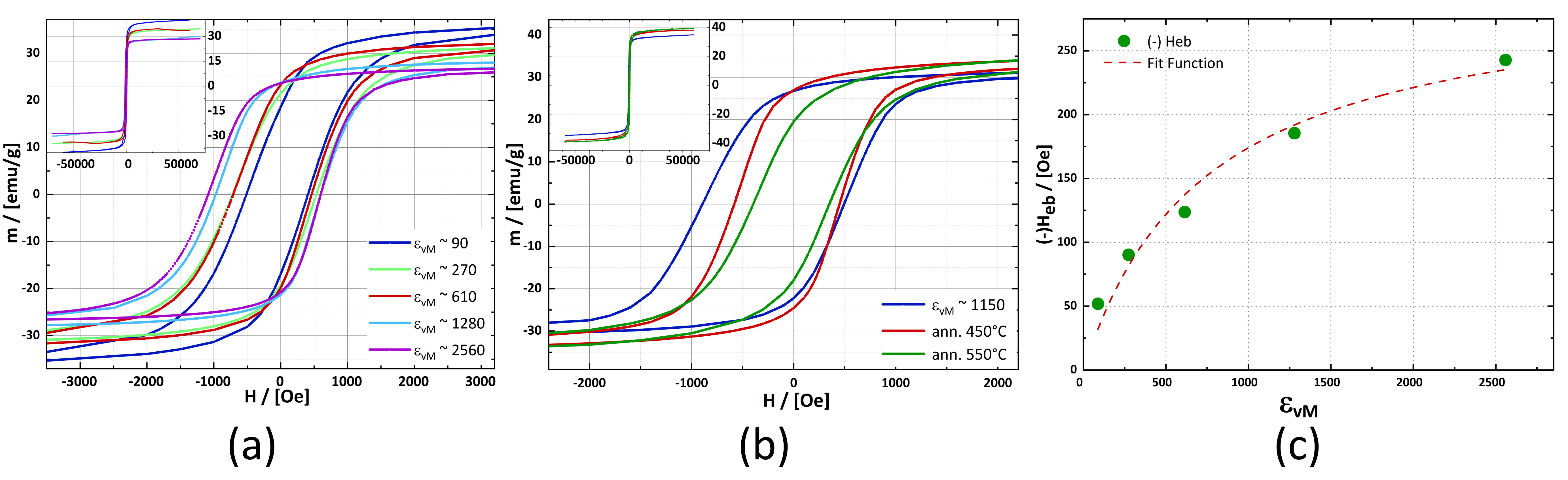}
\caption{The magnetic hysteresis loops of the Fe\textsubscript{10}Ni\textsubscript{40}NiO\textsubscript{50} sample were measured at 8 $K$. The graph displays a zoom-in at $\pm$ 2000 $Oe$. Inlet in the left corner provides the whole hysteresis loops. (\textbf{a}) Hysteresis loop of as-deformed samples (\textbf{b}) Hysteresis loops of the annealed samples. (\textbf{c}) H\textsubscript{eb} for different amounts of applied strain. The used fit is not based on a physical model and serves only to approximate the expected saturation of H\textsubscript{eb} for an infinite amount of applied strain. \label{fig6}}
\end{figure}

For SQUID investigations, the promising microstructures of the Fe\textsubscript{10}Ni\textsubscript{40}NiO\textsubscript{50} samples were chosen. The hysteresis loops shown in Figure~\ref{fig6}a have a continuously growing H\textsubscript{eb}, which could be correlated to the refinement of microstructure through the application of strain. The applied strain changed the hysteresis loop characteristic in general. As M\textsubscript{S} decreased, coercivity rose. The coercivity nearly doubled from $\epsilon\textsubscript{vM}$ $\sim$90 with H\textsubscript{C} = 460 $Oe$ to $\epsilon\textsubscript{vM}$ $\sim$2560 having H\textsubscript{C} = 842 $Oe$. Magnetic remanence (M\textsubscript{R}) rose from M\textsubscript{R} = 18.5 $emu/g$ to M\textsubscript{R} = 23.3 $emu/g$ at $\epsilon\textsubscript{vM}$ $\sim$610 and became independent of the applied strain for further deformation. H\textsubscript{eb} started at $\epsilon\textsubscript{vM}$ $\sim$90 with H\textsubscript{eb} = -52 $Oe$ and ended at $\epsilon\textsubscript{vM}$ $\sim$2560 with H\textsubscript{eb} = -243 $Oe$ (Table ~\ref{tab3}). As mentioned previously, M\textsubscript{S} reduced from M\textsubscript{S,}\textsubscript{$\epsilon\textsubscript{vM}$ $\sim$90} = 37.2 $emu/g$ to M\textsubscript{S,}\textsubscript{$\epsilon\textsubscript{vM}$ $\sim$2560} = 27.5 $emu/g$. These changes in M\textsubscript{S} and H\textsubscript{eb} appeared to be interrelated and might have had their origins in the detected changes in CSDS, enlargement of phase interfaces and the generation of $\gamma$-Fe\textsubscript{2}O\textsubscript{3}.  \\
The slight increase in Vickers microhardness detected at $r$ $\leq$1.25 $mm$ (Figure ~\ref{fig3}b) was therefore confirmed through the continuously growing H\textsubscript{eb} for such high amounts of applied strain and was in accordance with the impression of an ongoing refinement gained by SEM and XRD investigations. \\
Figure~\ref{fig6}b shows hysteresis loops of the annealed samples at 8 $K$. The annealing treatment caused a drastic reduction of H\textsubscript{eb} from H\textsubscript{eb,}\textsubscript{$\epsilon\textsubscript{vM}$ $\sim$1150} = -200 $Oe$ to H\textsubscript{eb,}\textsubscript{ann. 550°C} = -23 $Oe$ and M\textsubscript{R} (Table ~\ref{tab3}), whereas M\textsubscript{S} became larger, changing from M\textsubscript{S,}\textsubscript{$\epsilon\textsubscript{vM}$ $\sim$1150} = 32.3 $emu/g$ to M\textsubscript{S,}\textsubscript{ann. 550°C} = 36.6 $emu/g$. The rise of M\textsubscript{S} with annealing treatment correlated with the detected growth of CSDS for all involved phases (Table ~\ref{tab2}). \\
The annealing initiated topological changes of the phase interface morphology, leading to a reduction of interface roughness. This would be considered to affect the H\textsubscript{eb} as well \cite{Kumar2016,Nogues1999a}, but its impact would be minor compared to the growth of phase size, especially in a polycrystalline sample \cite {Nogues1999a}. The enlargement of FM-phase dimensions is qualitatively evident in the SEM micrographs shown in Figure ~\ref{fig5} and is presumed to have been the main reason for the decline of H\textsubscript{eb} by facilitating magnetic nucleation within the FM-phase, after annealing had been applied. \\
%%%
The detected change in H\textsubscript{eb} versus applied strain for the Fe\textsubscript{10}Ni\textsubscript{40}NiO\textsubscript{50} nanocomposite displayed a saturation behaviour similar to the microstructural saturation in single-phase systems previously reported for several phases deformed by HPT \cite{Pippan2006,Renk2019}. A simple fit model based on an asymptotic approximation yielded, for an asymptotic limit, an H\textsubscript{eb,}\textsubscript{max} of ~300 $Oe$ (Figure~\ref{fig6}c) for the Fe\textsubscript{10}Ni\textsubscript{40}NiO\textsubscript{50} composite. The fit model is not based on a physical theory and was only used here to approximate an assumed saturation of H\textsubscript{eb} through a saturation of microstructural refinement. \\
\begin{table}[H]
\caption{Summary of magnetic hysteresis loops of Fe\textsubscript{10}Ni\textsubscript{40}NiO\textsubscript{50} samples measured at 8 $K$, which are shown in Figure~\ref{fig6}a and b. Exchange bias (H\textsubscript{eb}), symmetric coercivity (H\textsubscript{C}), coercivity left side (H\textsubscript{C1}), coercivity right side (H\textsubscript{C2}), saturation magnetisation (M\textsubscript{S}) and magnetic remanence (M\textsubscript{R}). \label{tab3}}

		\newcolumntype{C}{>{\centering\arraybackslash}X}
		\begin{tabularx}{1\textwidth}{CCCCCCC}
			\hline\hline
			\empty	& \textbf{H\textsubscript{eb} / [$ Oe$] }	& \textbf{H\textsubscript{C} / [$ Oe$]}     & \textbf{H\textsubscript{C1} / [$ Oe$]} & \textbf{H\textsubscript{C2} / [$ Oe$]} & \textbf{M\textsubscript{S} / [$emu/g$]} & \textbf{M\textsubscript{R} / [$emu/g$]} \\
			\hline
$\epsilon$\textsubscript{vM}$\sim$90&	-52&	460&	-512&	408&	37.3&		18.6\\
$\epsilon$\textsubscript{vM}$\sim$270&	-90&		625&	-715&	535&32.6	&		21.2\\
$\epsilon$\textsubscript{vM}$\sim$610&	-124&	587&	-710&	463&	34.1&		23.3\\
$\epsilon$\textsubscript{vM}$\sim$1280&	-186&	786&	-971&	600&	28.8&		23.7\\
$\epsilon$\textsubscript{vM}$\sim$2560&	-243&	843&	-1085&	599&	27.6&		23.7\\
\hline
$\epsilon$\textsubscript{vM}$\sim$1150&	-200&	697&	-894&	494&	32.3&		26.8\\
ann. 450°C&	-67&	515&	-581&	449&	35.8&		27.0\\
ann. 550°C&	-23&	373&	-396&		349&	36.6&		19.6\\			             	 
			\hline\hline
		\end{tabularx}

	%\noindent{\footnotesize{* Tables may have a footer.}}
\end{table}
%\newpage
\section{Discussion}
\subsection{Deformation behaviour}
The deformation with HPT of composites containing '-ductile-brittle' phases has been reported for several phase systems including, Cu-Co and Cu-W \cite{Bachmaier2016,Sabirov2007}. For those systems, a fracturing of the '-brittle' phase has been proposed to have a crucial influence on the refinement and homogenisation of the sample's microstructure. Considering the Fe\textsubscript{50}NiO\textsubscript{50} sample at RT, such a fragmentation was not observed to an extensive degree. However, that does not imply that NiO cannot deform at RT. As shown in Figure ~\ref{fig1}a at RT the Fe and NiO deformed to lamella structures, but then reached a 'steady state' without fragmentation of the NiO phase in reasonably large amounts. This effect is clear when the SEM micrographs at $r$ $\sim$0 $mm$ and $r$ $\sim$3 $mm$ are compared (Figure ~\ref{fig1}a). In contrast, Vickers microhardness increased with radius for the sample deformed at RT, indicating an ongoing grain refinement within the phases (Figure ~\ref{fig1}c), which appeared to occur predominantly in the more ductile $\alpha$-Fe phase.\\
With a body-centred cubic crystal structure, $\alpha$-Fe has \{110\}<-111> as the primary slip system at RT. The \{110\}<-111> slip system family fulfils the von Mises criterion of five independent slip systems for the absorption of general strain via slip \cite{Groves1963}. The mechanical properties of $\alpha$-Fe are therefore different compared to ceramics such as NiO, which has an NaCl crystal structure with the primary slip systems \{110\}<1-10> at RT \cite{Groves1963}.  Basic consideration reveals, that such NaCl structures have two independent slip systems and that the von Mises criterion is not fulfilled; the absorption of plasticity by NiO is consequently limited. Hence, NiO cannot absorb general strain through slip, and the likelihood of the fragmentation of polycrystalline NiO is increased by the lack of matching slip systems of adjacent grains \cite{Groves1963}.\\
These considerations are important for better understanding of the deformation process in such complex systems. Whether the initial phase is $\alpha$-Fe or Ni, both phases are more ductile at RT than NiO. It could be that, especially at RT, the applied strain is absorbed by the $\alpha$-Fe to a large extent, as the activation of slip is assumed to be easier in $\alpha$-Fe than in the NiO-phase. It is plausible that through the elevated deformation temperatures, the activation of the additional slip systems is eased in NiO. The increased likelihood of slip in NiO could enhance fragmentation and consequently phase mixing. The eased activation of slip systems in NiO could be one reason for the drastic change of deformation behaviour at T\textsubscript{def} $\geq$225°C, which was seen for the Fe\textsubscript{50}NiO\textsubscript{50} samples.\\
The discrete change of deformation behaviour between 200°C and 225°C (Figure ~\ref{fig1}e), could not be related to a specific transition temperature. Although NiO has a T\textsubscript{N} of $\sim$251°C \cite{Srinivasan1984}, the expected change of plasticity around the T\textsubscript{N} was not reported in \cite{Guiberteau1986}. The brittle-to-ductile transition temperature (BDTT) of NiO has  been reported to be $\sim$0.3 of the melting temperature or approximately $\sim$600°C \cite{Evans1972}, which is still significantly above 225°C. In the same study, a brittle-to-ductile transition region was noted, ranging from 300°C to $\sim$550°C. Above 550°C, NiO is considered to be ductile due to the activation of additional slip systems \cite{Evans1972}.\\
A further important point is the slip transfer across phase boundaries, which was observed in \cite{Guetaz1994,Sinclair1999}. The likelihood for a glide or dislocation transfer between phases is higher, if the angle mismatch of adjacent slip systems is at a minimum \cite{Sinclair1999}, in addition to other conditions mentioned in \cite{Sinclair1999}. HPT processing at elevated temperatures could lower the activation energy of additional slip systems within NiO. Therefore, the likelihood of dislocation transfer between adjacent phases is increased, because more slip systems are available inside NiO, if processed at elevated temperatures. This assumption could also explain the abrupt change in deformation behaviour observed in the Fe\textsubscript{50}NiO\textsubscript{50} samples, when T\textsubscript{def} was elevated from T\textsubscript{def} = 200°C to T\textsubscript{def} = 225°C.\\
Furthermore, the possibility cannot be excluded, that the application of high hydrostatic pressure reduced the activation energy of additional slip systems in NiO and caused a shift to lower BDTT. Activation of additional slip systems through hydrostatic pressure has been discussed in the literature \cite{Wenk2000,Kang2009} and has been demonstrated for a Mg-alloy, in which non-basal glide systems were activated with an applied pressure of $\sim$125 $MPa$ \cite{Kang2009}. \\
Eased slip in NiO can lead to a roughening of interface morphology on a nanometre scale; this effect has been observed in other two-phase systems \cite{Guetaz1994}. The absorption of dislocations from local pile-ups, could cause local plastic instabilities, which are believed to support fragmentation of the more brittle phase, as discussed in a recent study \cite{Kormout2017}.\\
When the Fe\textsubscript{10}Ni\textsubscript{40}NiO\textsubscript{50} composition was deformed at T\textsubscript{def} = 300°C, the NiO-phase was conserved. In addition to the previously discussed impact of slip system activation and dislocation transfer in NiO, a work-hardening of the softer phase could also result in an enhanced fragmentation of NiO. Such a phenomenon has been reported for the Cu-Co-system during HPT-deformation \cite{Bachmaier2016}. The conservation of NiO facilitates a prolonged fragmentation, which could lead to the fine distribution of NiO imbedded in Ni (Figure ~\ref{fig3}a). \\
NiO can break off from larger agglomerates, which would ease the reduction by Fe through the creation of additional phase interfaces. It is likely that the accumulation of $\gamma$-Fe\textsubscript{2}O\textsubscript{3} occurs at phase interfaces or the grain boundary (GB). These small and dispersed clusters of $\gamma$-Fe\textsubscript{2}O\textsubscript{3} could inhibit GB-motion and accelerate grain refinement. In addition, through the efficient fragmentation of NiO, a highly diverse Ni and NiO-phase structure is created on a microscale, significantly hindering GB-motion and plastic deformation. Both phenomena could have led to the small observed CSDS of approximately 20 $nm$ according to the WH method at $\epsilon\textsubscript{vM}$ >1280 (Table ~\ref{tab2})  and could have caused the unusual high Vickers microhardness of 9 to 9.9 $GPa$.\\
The mismatch of slip systems and lattice parameters of participating phases, especially between $\gamma$-Fe\textsubscript{2}O\textsubscript{3} and the Ni- and NiO phases, could have led to a dislocation pile-up at the $\gamma$-Fe\textsubscript{2}O\textsubscript{3} phase, resulting in the creation of a non-crystalline region at the phase interface adjacent to $\gamma$-Fe\textsubscript{2}O\textsubscript{3}. Previous research has cited the build-up of a non-crystalline region at the phase interface of the Cu-Nb system as creating an impenetrable barrier for dislocation \cite{Raabe1995}.\\ 
Comparing the observed Vickers microhardness values in Figure ~\ref{fig3}b to the results obtained for nanocrystalline Ni \cite{Siow2004} or slightly oxidized Ni-powder processed by HPT \cite{Bachmaier2009} shows the influence of a nanostructured NiO phase on mechanical properties. Despite the reported high hardness values in of approximately 7 $GPa$ for nanocrystalline Ni or Ni with NiO clusters \cite{Siow2004,Bachmaier2009} the bulk nanocomposite of Fe\textsubscript{10}Ni\textsubscript{40}NiO\textsubscript{50} surpassed these results by 30-40\%.\\
%%%%
\subsection{Phase formation during deformation}
The efficient reduction of NiO and the observed formation of Fe\textsubscript{3}O\textsubscript{4} and $\gamma$-Fe\textsubscript{2}O\textsubscript{3} for both investigated compositions was unanticipated. Commercially available Ni powder is reported to oxidize at $>$350°C in ambient atmosphere \cite{Cabanas2012}. In general, NiO is considered to be a stable oxide, and therefore a reduction of Fe was not expected. This reduction can be explained by comparing the heat of formation, which is significantly less for NiO than for Fe\textsubscript{3}O\textsubscript{4} and $\gamma$-Fe\textsubscript{2}O\textsubscript{3} \cite{Cornell2003,Gaskell1996}. The difference in heat of formation could also explain the one-way characteristic of the reaction.\\
According to the data, the onset of NiO reduction by Fe during HPT processing was initiated between 200°C and 225°C (Figure ~\ref{fig2}). This reaction occurred much more efficiently at T\textsubscript{def} = 300°C in the Fe\textsubscript{50}NiO\textsubscript{50} material system. The formation to the thermodynamically more favourable Fe\textsubscript{3}O\textsubscript{4} appears to have been enhanced by the extreme conditions during sample synthesis combined with the fragmented and nanocrystalline NiO, which provided a very large amount of interfacial phase area for the observed reduction of NiO. A slip in the \{110\}<1-10> system in the NiO phase could facilitate the oxidation of Fe, because the occurrence of unsaturated oxygen bonds would be more likely. These two mentioned factors are considered to be the main cause for the observed two-phase system of Fe\textsubscript{3}O\textsubscript{4} and $\gamma$-Fe\textsubscript{39}Ni\textsubscript{61} for the Fe\textsubscript{50}NiO\textsubscript{50} material system, when deformed at T\textsubscript{def} = 300°C. \\
NiO could not be resolved in the XRD scan for the Fe\textsubscript{50}NiO\textsubscript{50} samples deformed at T\textsubscript{def} = 300°C (Figure ~\ref{fig2}), although 16-wt\% NiO should have been left after synthesis inside the nanocomposite. The possibility that some residual NiO remained after deformation is supported by the analysis of the $\gamma$-FeNi phase XRD-pattern. If complete reduction of NiO through Fe and complete incorporation of the residual Ni-atoms in the $\gamma$-FeNi phase is assumed, the composition of the $\gamma$-FeNi phase would change to $\gamma$-Fe\textsubscript{20}Ni\textsubscript{80}. The peak pattern of $\gamma$-Fe\textsubscript{20}Ni\textsubscript{80} had its main peak at  q\textsubscript{z} = 3.073  \AA\textsuperscript{-1} and was therefore distinguishable from the peak position of the observed $\gamma$-Fe\textsubscript{39}Ni\textsubscript{61} phase. \\
The XRD results of Fe\textsubscript{50}NiO\textsubscript{50} and Fe\textsubscript{10}Ni\textsubscript{40}NiO\textsubscript{50} samples in Figures ~\ref{fig2} and ~\ref{fig4}, respectively, detected different formations of the Fe\textsubscript{x}O\textsubscript{y} phase through the reduction of NiO. The formation of Fe\textsubscript{x}O\textsubscript{y} is influenced by the amount of available O \cite{Wriedt1991}. In the Fe\textsubscript{10}Ni\textsubscript{40}NiO\textsubscript{50} sample, the Fe content was much lower compared to Fe\textsubscript{50}NiO\textsubscript{50}, and therefore sufficient O was available to form $\gamma$-Fe\textsubscript{2}O\textsubscript{3}. \\
Moreover, $\gamma$-Fe\textsubscript{2}O\textsubscript{3} is thermodynamically more stable than Fe\textsubscript{3}O\textsubscript{4} \cite{Cornell2003} and therefore is considered more likely to be formed, provided that sufficient O is available during deformation. Another difference from the Fe\textsubscript{50}NiO\textsubscript{50} samples at T\textsubscript{def} = 300°C is the neglectable substitution of Fe inside the Ni. After the deformation of Fe\textsubscript{10}Ni\textsubscript{40}NiO\textsubscript{50} composition, WAXS detected a $\gamma$-phase consisting solely of Ni. \\
\subsection{Magnetic properties}
Regarding the magnetometry results, the aforementioned Fe-oxide phases (i.e. Fe\textsubscript{3}O\textsubscript{4} and $\gamma$-Fe\textsubscript{2}O\textsubscript{3}) were both FM, but distinguished themselves in their magnetic properties, for example bulk-M\textsubscript{S,}\textsubscript{Fe\textsubscript{3}O\textsubscript{4}} = 98.6 $emu/g$ \cite{Caruntu2007} and bulk-M\textsubscript{S,}\textsubscript{$\gamma$-Fe\textsubscript{2}O\textsubscript{3}} = 85.6 $emu/g$ \cite{Roca2007} at 8 $K$. Based on the results from WAXS measurements, the M\textsubscript{S,}\textsubscript{$\gamma$-Fe\textsubscript{2}O\textsubscript{3}} is used in the following discussion of the results for Fe\textsubscript{10}Ni\textsubscript{40}NiO\textsubscript{50} nanocomposite. \\
\begin{figure}[H]
\center
\includegraphics[width=9 cm]{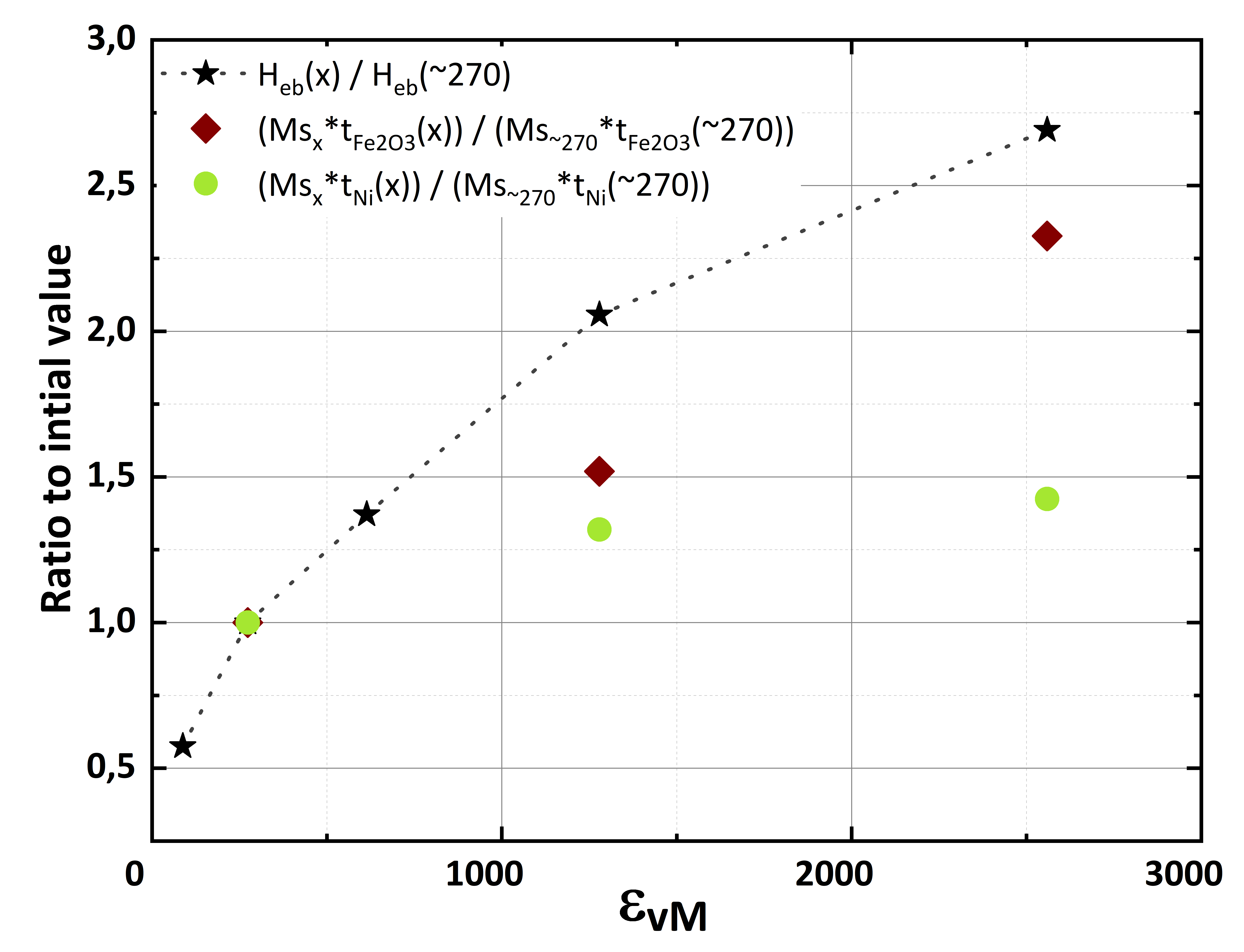}
\caption{Relative change of H\textsubscript{eb} related to H\textsubscript{eb,}\textsubscript{$\epsilon\textsubscript{vM}$ $\sim$270} and comparison of the ratio to the relative changes of M\textsubscript{S} and CSDS size of t\textsubscript{$\gamma$-Fe2O3} and t\textsubscript{Ni}. \label{fig7}}
\end{figure}      
The H\textsubscript{eb} can be approximated for thin films with the simple theoretical model of \(H\textsubscript{eb} \sim \sigma/(M\textsubscript{S} \cdot t\textsubscript{FM}) \)($\sigma$ = interfacial coupling energy density, t\textsubscript{FM} = FM-thickness) \cite{Berkowitz1999}. Utilising this model, with the assumption that it is valid to some extent for nanocomposites measured at $8K$ as well, allows estimation of the origin of the observed change of H\textsubscript{eb} with applied strain.\\
The increase of H\textsubscript{eb,}\textsubscript{$\epsilon\textsubscript{vM}$$\sim$90} = -52 $Oe$ to H\textsubscript{eb,}\textsubscript{$\epsilon\textsubscript{vM}$$\sim$2560} = -242 $Oe$ cannot be explained by the decline of M\textsubscript{S} alone. As shown in Table ~\ref{tab3}, M\textsubscript{S} decreased by approximately 26\% from M\textsubscript{S,}\textsubscript{$\epsilon\textsubscript{vM}$$\sim$90} = 37.3 $emu/g$ to M\textsubscript{S,}\textsubscript{$\epsilon\textsubscript{vM}$$\sim$2560} = 27.6 $emu/g$, whereas H\textsubscript{eb} increased almost fivefold from H\textsubscript{eb,}\textsubscript{min} to H\textsubscript{eb,}\textsubscript{max}. Considering the observed changes of CSDS (Table ~\ref{tab2}) in addition to the decrease of M\textsubscript{S}, yields a more accurate estimation (Figure~\ref{fig7}). The increase of H\textsubscript{eb,}\textsubscript{$\epsilon\textsubscript{vM}$$\sim$270} = -90 $Oe$ to H\textsubscript{eb,}\textsubscript{$\epsilon\textsubscript{vM}$$\sim$2560} = -242 $Oe$ can be explained to a great extent by the combined decreases of M\textsubscript{S} and CSDS of $\gamma$-Fe\textsubscript{2}O\textsubscript{3} using the approximation that $\sigma$ was the same for both applied strains. This explanation implies, that NiO was exchange biased to the $\gamma$-Fe\textsubscript{2}O\textsubscript{3} as well as to the Ni-crystallites. From this calculation it can be deduced that $\gamma$-Fe\textsubscript{2}O\textsubscript{3} crystallites dominated the magnetic characteristic of the Fe\textsubscript{10}Ni\textsubscript{40}NiO\textsubscript{50} samples and that the Ni-crystallites contributed partially.\\
The annealing experiments allow a deeper insight into the dependency of H\textsubscript{eb} on the nanocomposite microstructure. The variable M\textsubscript{S} recovered after deformation, from M\textsubscript{S,}\textsubscript{$\epsilon\textsubscript{vM}$ $\sim$1150} = 32.3 $emu/g$ to M\textsubscript{S,}\textsubscript{ann. 550°C} = 36.9 $emu/g$, and H\textsubscript{eb} decreased to $\sim$10\% of its initial value of H\textsubscript{eb,}\textsubscript{as def} = -200  $Oe$ through annealing at 550°C. The annealed samples exhibited continuous CSDS growth for all three phases (Table ~\ref{tab2}) as well as phase size growth (Figure ~\ref{fig5}). Changes in CSDS in $\gamma$-Fe\textsubscript{2}O\textsubscript{3} were approximately twofold as well, but the decrease of H\textsubscript{eb} was more significant and could be described only partially by the previous estimation.\\
Although CSDS apparently had a crucial influence on H\textsubscript{eb}, magnetic properties in annealed samples can be influenced by other effects as well. The application of temperature can initialize optimisation of phase interface area to lower surface energies and cause a reduction of phase interface topology. Those effects should also transform the morphology of the FM-AFM interface to a sharper phase gradient. The decreased interfacial area could affect the enclosure of FM-grains by the AFM phase and reduce the overall magnetic stiffness inside the FM-phase. Phase size growth could enable magnetic nucleation during field reversal, which could further weaken H\textsubscript{eb}.\\
The variation of M\textsubscript{S} with applied strain or annealing temperature appears to have been related to some degree to the rise of CSDS of $\gamma$-Fe\textsubscript{2}O\textsubscript{3}, due to its large relative changes in size. Previous studies have found that FM-nanocrystallites possess a size-dependent M\textsubscript{S} in general. The decline of M\textsubscript{S} is related to a non-magnetic region at the crystallite boundary, which begins to influence M\textsubscript{S} non-linearly below approximately 20 $nm$ of crystallite dimension through a growing surface-to-volume ratio \cite{Berkowitz1968}.\\
\begin{figure}[H]
\center
\includegraphics[width=9 cm]{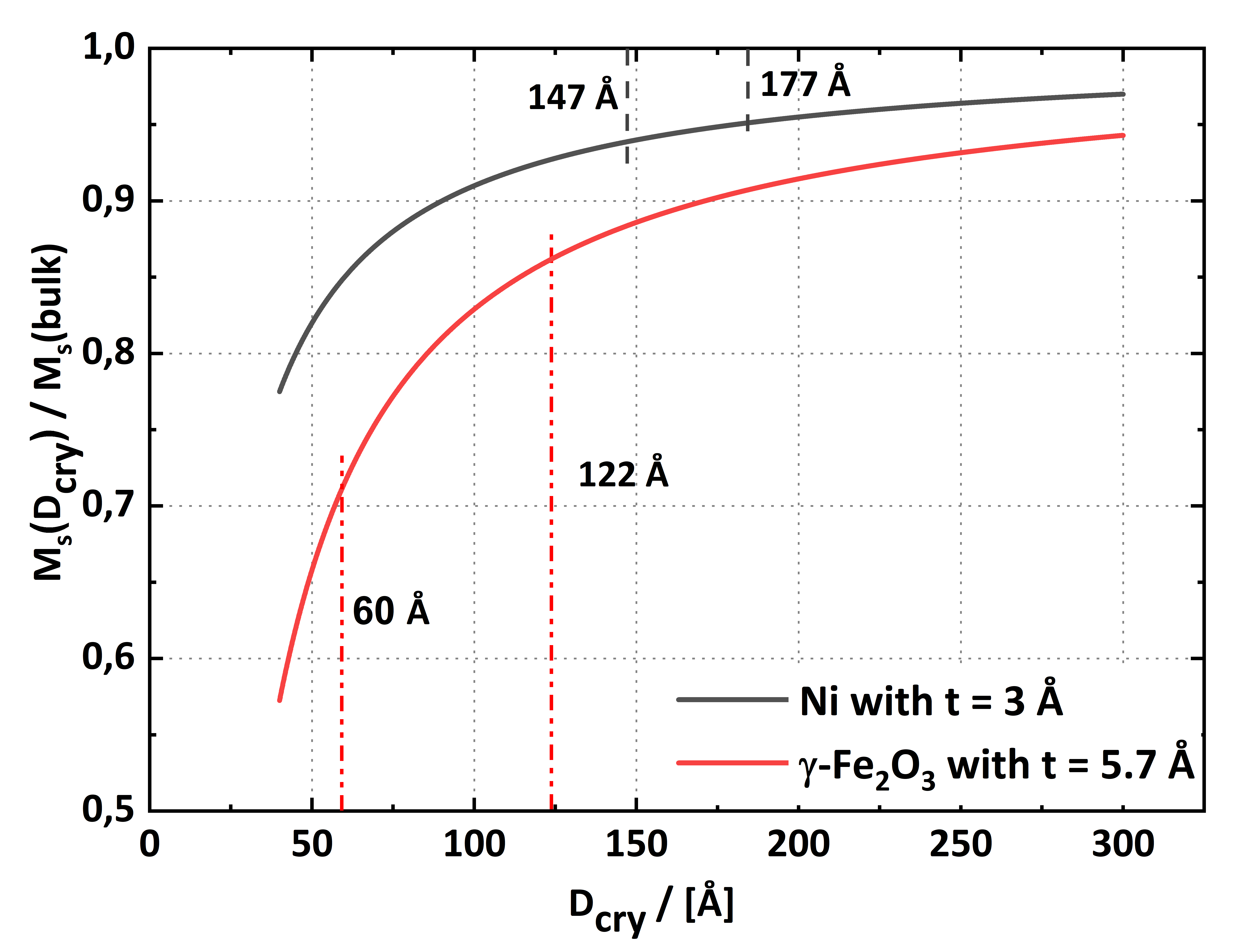}
\caption{The influence of crystallite size on M\textsubscript{S}, according to \cite{Berkowitz1968}, assuming a non-magnetic layer between adjacent crystallites of 5.7 \AA\hspace{0.5mm}  for $\gamma$-Fe\textsubscript{2}O\textsubscript{3} and 3 \AA\hspace{0.5mm} for Ni. Due to the smaller CSDS for $\gamma$-Fe\textsubscript{2}O\textsubscript{3} the surface-to-volume ratio becomes larger, causing a stronger decrease of M\textsubscript{S}\textsubscript{$\gamma$-Fe\textsubscript{2}O\textsubscript{3}}. \label{fig8}}
\end{figure}   
To obtain a better insight, the simple approximation from \cite{Berkowitz1968} was used for M\textsubscript{S}, based on the relationship $M\textsubscript{S}(D\textsubscript{Cry}) = M\textsubscript{S}(bulk) \cdot (1- 3\cdot t\textsubscript{non-mag}/D\textsubscript{Cry})$(D\textsubscript{Cry}...crystallite size; for the calculation, the CSDSs from Table ~\ref{tab2} and t\textsubscript{non-mag}...thickness of a non-magnetic layer between adjacent nanocrystallites were used) for nanocrystalline material. Utilising this model for the Fe\textsubscript{10}Ni\textsubscript{40}NiO\textsubscript{50} nanocomposite, which consisted of 12-wt\% $\gamma$-Fe\textsubscript{2}O\textsubscript{3} and 48.6-wt\% Ni; yields with a non-magnetic layer of 5.7 \AA\hspace{0.5mm}  for $\gamma$-Fe\textsubscript{2}O\textsubscript{3} \cite{Berkowitz1968}, 3 \AA\hspace{0.5mm}  for Ni (estimated for Ni) and assuming a bulk M\textsubscript{S,}\textsubscript{Ni} = 58.6 $emu/g$  \cite{Crangle1971} and M\textsubscript{S,}\textsubscript{$\gamma$-Fe\textsubscript{2}O\textsubscript{3}} = 85.6 $emu/g$\cite{Roca2007}:
M\textsubscript{cal.S,}\textsubscript{$\epsilon\textsubscript{vM}$$\sim$270} = 35.7 $emu/g$, M\textsubscript{cal.S,}\textsubscript{$\epsilon\textsubscript{vM}$$\sim$1280} = 35.1 $emu/g$ and M\textsubscript{cal.S,}\textsubscript{$\epsilon\textsubscript{vM}$$\sim$2560} = 34.2 $emu/g$, far more than measured (Table ~\ref{tab3}). These results suggest that the assumed thickness of the non-magnetic layer was underestimated for the considered nanocomposite and should have been much higher to match the experimental results. Regarding the different contributions to M\textsubscript{S} of $\gamma$-Fe\textsubscript{2}O\textsubscript{3} and Ni, crystallites that possess different CSDS allow the conclusion, that the decrease of M\textsubscript{S} was mainly caused by the shrinking CSDS of $\gamma$-Fe\textsubscript{2}O\textsubscript{3} crystallites through the deformation (Figure ~\ref{fig8}).\\
The results from the annealed samples support the previous argument, that the $\gamma$-Fe\textsubscript{2}O\textsubscript{3} phase mainly influenced M\textsubscript{S}. Growth of $\gamma$-Fe\textsubscript{2}O\textsubscript{3} from 8 $nm$ to 14 $nm$ CSDS would lead to the following calculated M\textsubscript{S}: M\textsubscript{cal.S,}\textsubscript{$\epsilon\textsubscript{vM}$ $\sim$1150} = 34.4 $emu/g$, M\textsubscript{cal.S,}\textsubscript{ann.450°C} = 35.0 $emu/g$ and M\textsubscript{cal.S,}\textsubscript{ann.550°C} = 36.0 $emu/g$. The estimation for M\textsubscript{S,}\textsubscript{$\epsilon\textsubscript{vM}$ $\sim$1150} is certainly higher than the measured M\textsubscript{S} value; however, M\textsubscript{S,}\textsubscript{ann.450°C} and M\textsubscript{S,}\textsubscript{ann.550°C} are in good accordance with the values presented in Table ~\ref{tab3}. \\
The discrepancy between M\textsubscript{S,}\textsubscript{$\epsilon\textsubscript{vM}$ $\sim$1150} and M\textsubscript{S,}\textsubscript{$\epsilon\textsubscript{vM}$$\sim$2560} could be rooted in a larger number of non-magnetic regions than assumed. If  complete oxidation of Fe to $\gamma$-Fe\textsubscript{2}O\textsubscript{3} is assumed and crystal size effects are neglected, the calculated M\textsubscript{S} for the annealed Fe\textsubscript{10}Ni\textsubscript{40}NiO\textsubscript{50} sample yields, a value of M\textsubscript{S,}\textsubscript{cal} = 38.3 $emu/g$. The M\textsubscript{S,}\textsubscript{ann. 550°C} = 36.9 $emu/g$ is strikingly similar to the theoretical calculation. This comparison suggests that the non-magnetic phase was predominantly composed of Ni and $\gamma$-Fe\textsubscript{2}O\textsubscript{3} and was reduced to a minimum by the annealing treatment.\\ 
There have been reports of a phase transition from FM $\gamma$-Fe\textsubscript{2}O\textsubscript{3} to the thermodynamically more favourable AFM $\alpha$-Fe\textsubscript{2}O\textsubscript{3} at temperatures of 370-600°C \cite{Cornell2003}. While this transition could have occurred during annealing, an accumulation of non-crystalline regions through the application of strain is more likely. On one hand, the XRD peak pattern of $\alpha$-Fe\textsubscript{2}O\textsubscript{3} is distinct from $\gamma$-Fe\textsubscript{2}O\textsubscript{3};  on other hand, a formation of $\alpha$-Fe\textsubscript{2}O\textsubscript{3} cannot explain the rise in M\textsubscript{S} through annealing. The plausible existence of non-crystalline regions at interfaces, presumably containing $\gamma$-Fe\textsubscript{2}O\textsubscript{3}, would have a fraction of bulk M\textsubscript{S}, M\textsubscript{S}\textsubscript{,amorphous} $\sim$2-5\% \cite{Machala2007}, and the transition from crystalline $\gamma$-Fe\textsubscript{2}O\textsubscript{3} phase to a phase with non-crystalline regions at the crystallite boundary could explain the changes in M\textsubscript{S}, detected when large amounts of strain had been applied. Composites, such as Nb-Cu, deformed by SPD have been reported to contain non-crystalline regions at phase interfaces \cite{Raabe1995,Sauvage2001}. Crystallisation of these non-crystalline regions was observed when annealing was applied \cite{Raabe1995}. These results suggest that similar behaviour occurred in the Fe\textsubscript{10}Ni\textsubscript{40}NiO\textsubscript{50} nanocomposite.\\
%%
%%%%%%%%%%%%%%%%%%%%%%%%%%%%%%%%%%%%%%%%%%
\section{Conclusions}
This study presents an insight into the deformation behaviour of phases possessing very distinct mechanical properties. Although NiO has, due to its NaCl-structure, different primary slip systems compared to Fe or FeNi, it was possible to synthesise nanocomposites with a homogeneous microstructure on the sub-micrometre regime. The importance of thermal HPT processing was demonstrated on the Fe\textsubscript{50}NiO\textsubscript{50} composition. Through the deformation at 300°C, a homogeneous two-phase microstructure evolved. XRD investigations confirmed a reduction of NiO through Fe leading to the formation of Fe\textsubscript{3}O\textsubscript{4} and $\gamma$-Fe\textsubscript{39}Ni\textsubscript{61}. \\
To conserve the AFM NiO in respect to enhance the H\textsubscript{eb}, a composition of Fe\textsubscript{10}Ni\textsubscript{40}NiO\textsubscript{50} was synthesised by HPT at 300°C and nanocomposites with a unique microstructure were obtained. SEM micrographs showed a very homogeneous microstructure, which primarily contained of NiO and Ni according to WAXS investigations. The formation of an additional phase was detected by WAXS, which consisted of $\gamma$-Fe\textsubscript{2}O\textsubscript{3}. Vickers microhardness measurements detected a microstructural saturation behaviour with a slight increase to outer radii to an unusual high Vickers microhardness of $\sim$9.9 $GPa$. WAXS, SEM and magnetometry results conveyed an ongoing refinement of the nanocomposite's microstructure. \\
It was possible to demonstrate the existence of H\textsubscript{eb} in such bulk sized nanocomposite and the continual increase of H\textsubscript{eb} with applied strain was correlated to the evolving microstructure. The rise of H\textsubscript{eb} was mainly attributed to a simultaneous decrease in M\textsubscript{S}, reduction of FM-phase size dimensions and crystallite size of the FM-phases, predominantly $\gamma$-Fe\textsubscript{2}O\textsubscript{3}. \\
Annealing experiments of Fe\textsubscript{10}Ni\textsubscript{40}NiO\textsubscript{50} samples showed a decrease of H\textsubscript{eb} and an increase of M\textsubscript{S}. Those changes of H\textsubscript{eb} and M\textsubscript{S} were contributed to the growth of FM-phase dimensions and the subsequent reduction of phase interfaces. \\
\\
\\
\section[\appendixname~\thesection]{Appendix A. Schematic Illustration of the HPT - Synthesis}
%All appendix sections must be cited in the main text. In the appendices, Figures, Tables, etc. should be labeled, starting with ``A''---e.g., Figure A1, Figure A2, etc.
\begin{figure}[H]
\center
\includegraphics[width=15 cm]{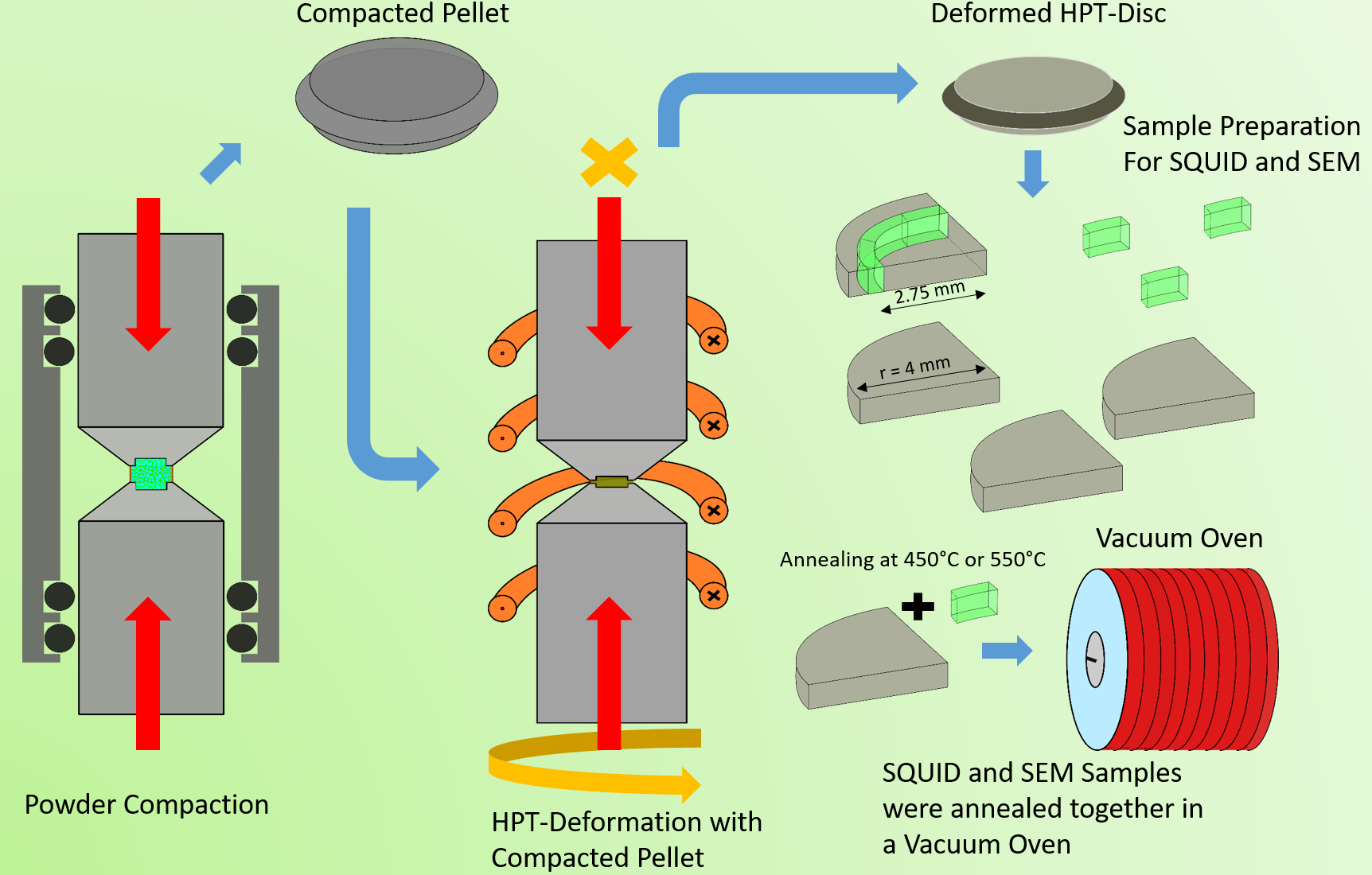}
\caption{Schematic description of the sample synthesis. From left to right: Powder blends were compacted with HPT inside an airtight capsule. The obtained pellet was used for HPT-deformation experiments. An inductive heating system provided the needed processes temperature for the HPT-anvils. HPT-disc was cut into the desired size for further investigations. The SQUID samples were cut out at the desired radius. For the annealing experiments a HPT-disc was cut into quarters. One quarter provided the SQUID-samples, which were cut out at the same radius. The SQUID sample was simultaneously annealed with a quarter of the HPT-disc. \label{figA1}}
\end{figure}
\textbf{Funding:}\\
This project has received funding from the European Research Council (ERC) under the European Union’s Horizon 2020 research and innovation programme (grant agreement No:757333). \\
\\
\textbf{Conflicts of Interest:}\\
The authors declare no conflict of interest.\\
\\
\textbf{Acknowledgement:}\\
We acknowledge DESY (Hamburg, Germany), a member of the Helmholtz Association HGF, for the provision of experimental facilities. Parts of this research were carried out at (PETRA III) and we would like to thank (N. Schell and E. Maawad) for assistance in using (P07B- High Energy Materials Science).

\bibliographystyle{abbrv}%elsarticle-harv %elsarticle-num-names %elsarticle-num (geht nicht-usepackages collide) %abbrvnat
\bibliography{Zawodzki_ExchangeBias_vs_Strain}

\end{document}